\documentclass[oldversion]{aa} 
\usepackage{graphicx}
\usepackage{color}
\usepackage{longtable}
\usepackage{txfonts}
\usepackage{rotating}
\usepackage{lscape}
\usepackage{natbib}
\newcommand{\gsim}{\;\lower.6ex\hbox{$\sim$}\kern-7.75pt\raise.65ex\hbox{$>$}\;}
\newcommand{\lsim}{\;\lower.6ex\hbox{$\sim$}\kern-7.75pt\raise.65ex\hbox{$<$}\;}

\begin{document}

\title{NGC~6535: the lowest mass Milky Way globular cluster with a Na-O 
anti-correlation?\thanks{Based on  observations
collected at ESO telescopes under  programme 093.B-0583}\fnmsep\thanks{
   Table 2 is only available in electronic form at the CDS via anonymous
   ftp to {\tt cdsarc.u-strasbg.fr} (130.79.128.5) or via
   {\tt http://cdsweb.u-strasbg.fr/cgi-bin/qcat?J/A+A/???/???}}
 }
 \subtitle{Cluster mass and age in the multiple population context.}

\author{
A. Bragaglia\inst{1}, 
E. Carretta\inst{1},
V. D'Orazi\inst{2,3,4},
A. Sollima\inst{1},
P. Donati\inst{1},
R.G. Gratton\inst{2},
\and
S. Lucatello\inst{2}
}

\authorrunning{A. Bragaglia et al.}
\titlerunning{Abundance analysis in NGC 6535}

\offprints{A. Bragaglia, angela.bragaglia@oabo.inaf.it}

\institute{
INAF-Osservatorio Astronomico di Bologna, via Gobetti 93/3, I-40129 Bologna, Italy
\and
INAF-Osservatorio Astronomico di Padova, vicolo dell'Osservatorio 5, I-35122
 Padova, Italy
\and
Monash Centre for Astrophysics, School of Physics and Astronomy, Monash University, Melbourne, VIC 3800, Australia
\and
Department of Physics and Astronomy, Macquarie University, Sydney, NSW 2109, Australia
  }

\date{}

\abstract{To understand globular clusters (GCs) we need to comprehend how their
formation process was able to produce their abundance distribution of light
elements. In particular, we seek to figure out which stars imprinted the
peculiar chemical signature of GCs. One of the best ways is to study the
light-element anti-correlations in a large sample of GCs that are analysed
homogeneously. As part of our spectroscopic survey of GCs with FLAMES, we
present here the results of our study of about 30 red giant member stars in the
low-mass, low-metallicity Milky Way cluster NGC~6535. We measured the metallicity (finding [Fe/H]=$-1.95$, rms=0.04 dex in our homogeneous scale) and
other elements of the cluster and, in particular, we concentrate here on O and Na abundances. These
elements define the normal Na-O anti-correlation of classical GCs, making
NGC~6535  perhaps the lowest mass cluster with a confirmed presence of multiple
populations. We updated the census of Galactic and extragalactic GCs for which a
statement on the presence or absence of multiple populations can be made on the
basis of high-resolution spectroscopy preferentially, or photometry and
low-resolution spectroscopy otherwise; we also discuss the importance of mass and age of the clusters as factors for multiple populations. }

\maketitle

\begin{figure}
\centering
\includegraphics[scale=0.5,bb=40 160 550 640,clip]{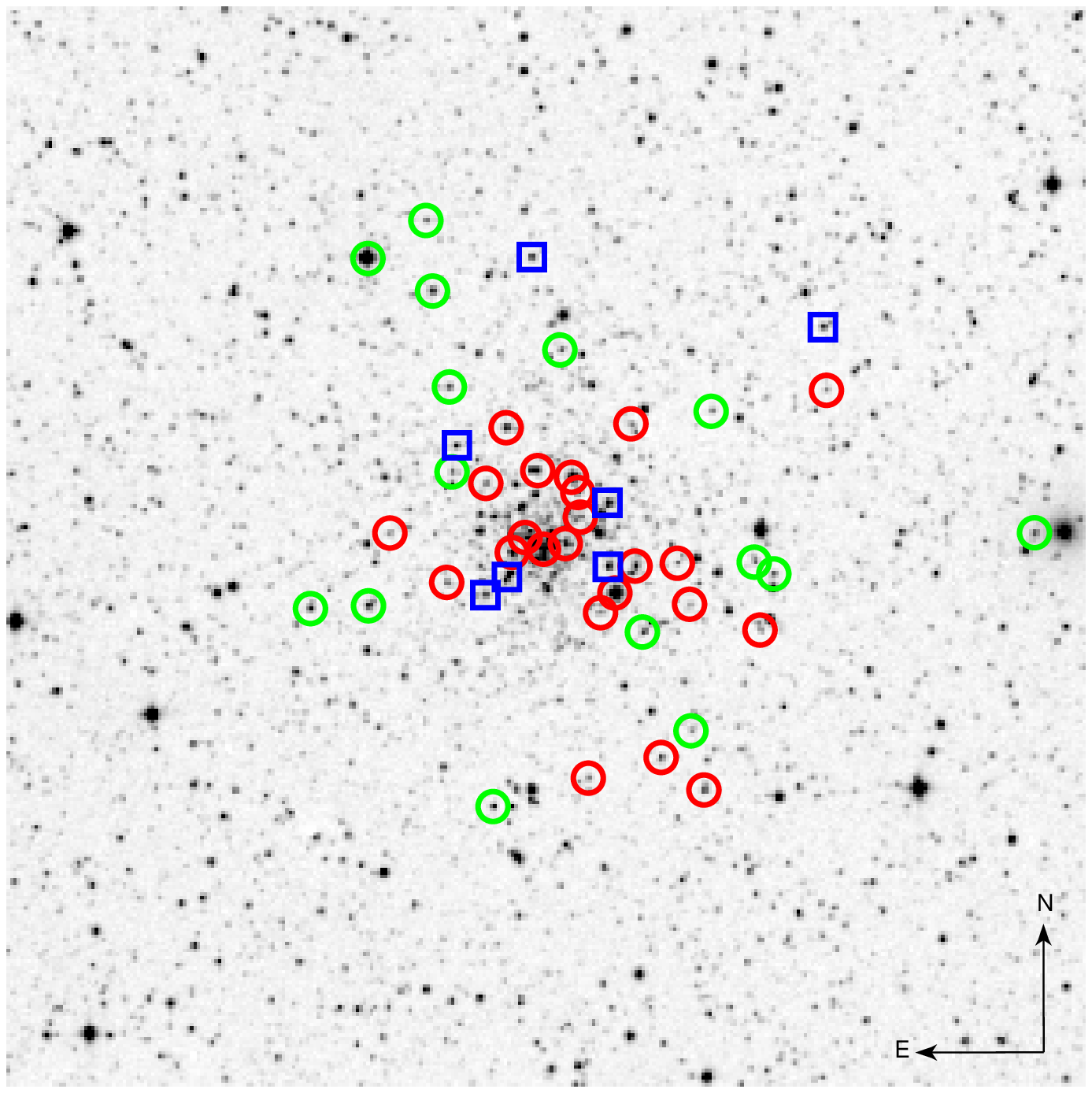}
\includegraphics[scale=0.43,bb=30 150 565 720,clip]{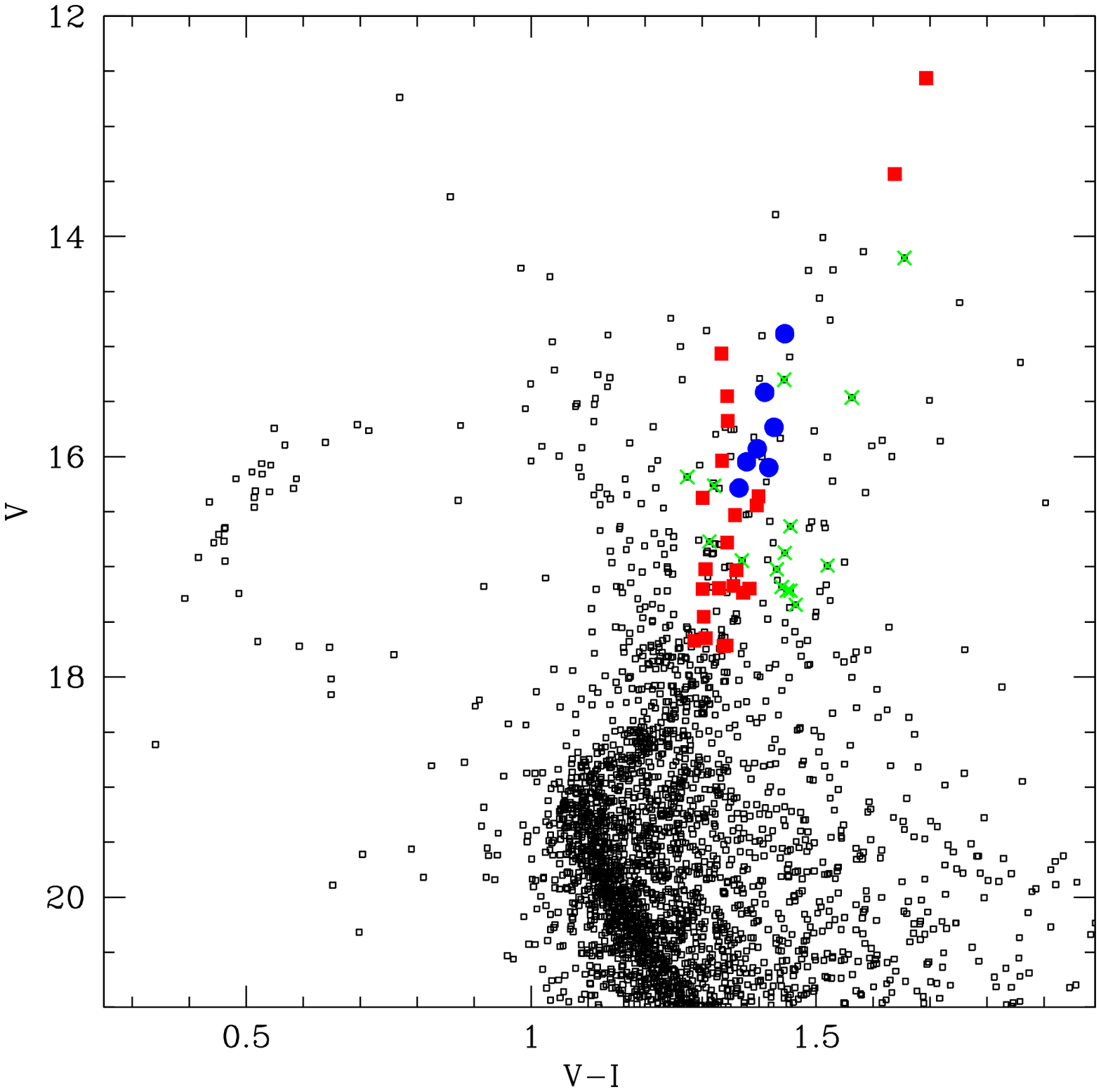}
\caption{Upper panel: Region of $12\arcmin \times 12\arcmin$ with the stars
observed indicated by coloured open  symbols (blue boxes: UVES; red circles:
GIRAFFE, members; green circles: GIRAFFE, non-members).  Lower panel: CMD of
NGC~6535 (photometry by \citealt{testa01}) is shown; our targets are indicated by larger
symbols with colours as above.}
\label{cmdoss}
\end{figure}

\section{Introduction}\label{intro}

Once considered as a good example of simple stellar populations (SSP), Galactic
globular clusters (GCs)  are currently thought to have formed in a complex chain
of events, which left a fossil record in their chemical composition, in
particular in their light elements He, C, N, O, Mg, Al, and Na    \citep[see e.g.
the review by][]{gratton12}.  Spectroscopically, almost all Milky Way (MW) GCs
studied host multiple stellar populations that can be traced by the
anti-correlated  variations of light elements C and N
\citep[e.g.][]{kayser08,smolinsky11}, O and Na, and Mg and Al (see e.g. 
\citealt{carretta09a,carretta09b,bragaglia6139,n5634} and references therein
for  our FLAMES survey of more than 25 MW GCs). Photometrically, light element
variations manifest in colour-magnitude diagrams (CMD) as sequence broadening
and splitting \citep[e.g.][and references therein]{stromgren,milone12,sumo,lee15,piottouv} which are particularly evident
when appropriate combinations of UV filters (tracing CN, OH, NH  bands) are
used.

In normal, massive GCs at least two populations coexist. One has a composition
that is indistinguishable from field stars of similar metallicity and is believed to be
the long-lived remnant of the first generation (FG) formed in the cluster. The
other, with a modified composition, is the second generation (SG), polluted by
the most massive stars of the FG with ejecta from H burning at high temperature
\citep{dd89,langer}.  Unfortunately, the question of which FG stars produced the
gas of modified composition is still unsettled (see e.g.
\cite{ventura01,decressin07,demink09,mz12,dh14}, and \cite{bastian15}). Hence we
still do not fully understand how GCs, and their multiple populations, (MPs)
formed.  To attack the problem, we need to combine spectroscopic, photometric,
and astrometric observations (yielding abundances, kinematics, and their spatial
distribution in GCs) with theoretical modelling (stellar evolution, formation,
and chemo-dynamical evolution of clusters).

On the observational front, it is crucial to study clusters covering the widest
range of properties, that is mass, metallicity, age, structural parameters, and
environment. While almost all MW GCs studied thus far show MPs, there are a few
exceptions, such as Ruprecht~106 \citep{villanovarup106}, Palomar~12
\citep{cohenpal12}, and Terzan~7 \citep{sbordoneter7}. Interestingly, the last
two are also associated with the disrupting Sagittarius dwarf galaxy (Sgr dSph),
the closest extragalactic environment. As of today, MPs have been detected in
several extragalactic GCs. High-resolution spectroscopy has been used for old,
massive clusters of the Magellanic Clouds (MCs); see
\cite{johnson06,mucciarelli09} for the Large Magellanic Cloud (LMC), and
\cite{dalessandro16} for the Small Magellanic Cloud (SMC), or the Fornax dwarf
spheroidal \citep[Fnx dSph;][]{letarte06}. Low-resolution spectroscopy and
photometry have been employed to study younger clusters in the MCs or the Fnx
dSph; see e.g. \cite{hollyhead17,niederhofer17b,larsen14} discussed in
Sect.~\ref{massage}. While these clusters can be used to explore a possible
dependence on age and environment of the MPs, these clusters are comparable in mass to the
bulk of the MW GCs and do not allow us to evaluate the impact of the environment on
the low-mass end of the GC mass distribution. 

We found that, apparently, there is an observed minimum cluster mass for
appearance of the  Na-O anti-correlation, i.e. of SG (see \citealt{zcorri}, 
Fig.~1 in \citealt{be39}, and Sect.~\ref{massage} for further discussion). This
is an important constraint for GC formation mechanisms because it indicates the
mass at which we expect that a GC is able to retain part of the ejecta of the
FG.  It is important to understand if this limit is real or due to the small
statistics, since only a handful of low-mass clusters have been studied and
only a few stars in each were spectroscopically observed.  The problem of
statistics can be bypassed using low-resolution spectroscopy and photometry,
which can reach fainter magnitudes and  larger samples than high-resolution
spectroscopy; however, this means that only C and N--- plus O with Hubble Space Telescope (HST) far-UV
filters, and possibly He--- can be studied. To increase the sample studied with
high-resolution spectroscopy, we obtained data for a few old and massive
open clusters \citep[OCs; i.e. Berkeley~39, NGC~6791:][]{be39,bragaglia6791}
and low-mass GCs, which are also connected with the Sgr dSph \citep[Terzan~8, NGC~6139,
and NGC~5634;][]{ter8,bragaglia6139,n5634}. 

We concentrate here on the low-mass GC  NGC~6535, which has $M_V=-4.75$
\citep[][and 2010 web update]{harris}. This would be the lowest present-day mass
GC in which an Na-O anti-correlation is found, since the present record holder,
Palomar 5, has $M_V=-5.17  $,  after shedding a good fraction of its mass, as
witnessed by its tidal tails (e.g. \citealt{odenkirchen01}). In Section 2 we
present literature information on the cluster, in Section 3 we describe the photometric data,  spectroscopic observations, and derivation of
atmospheric parameters. The abundance analysis is presented in Section 4 and a
discussion on the light-element abundances is given in Section 5. Kinematics is
discussed in  Section 6 and the age and mass limits for the appearance of
multiple populations in Section 7. A summary and conclusion are given in Section
8.

\section{NGC~6535 in literature}\label{lit} 

NGC~6535 is a low-concentration, low-mass GC; its absolute visual magnitude, a
proxy for present-day mass, is $M_V=-4.75$, the King-model concentration is
$c=1.33$, and the core radius and half-mass radius are $r_c=0.36\arcmin$,
$r_h=0.85\arcmin$  \citep[][2010 web update]{harris}. NGC 6535 is located  towards the
centre of the MW, at $l= 27.18\degr$, $b= 10.44\degr$, and suffers from severe
field contamination (see Fig.~\ref{cmdoss}). Notwithstanding its present small
Galactocentric distance, the cluster is metal-poor (R$_{GC}=3.9$ kpc,
[Fe/H]=-1.79; from \citealt{harris}) and is considered a halo GC. While there
are several photometric papers in the literature, it has never been studied with
high-resolution spectroscopy before.

After the first colour-magnitude diagram (CMD) obtained by \cite{liller}, the
cluster was studied by \cite{att} and \cite{sarajedini}.  
\cite{rosenberg99,rosenberg00} used the $V,V-I$ ground-based  CCD data of this
and more than 30 other GCs  to determine its age and found it coeval with the
bulk of GCs.  \cite{marin-franch}, using data obtained within the Hubble Space
Telescope {\em ACS Treasury Program} on GCs \citep{treasury}, placed NGC~6535 among
the young clusters.  An old absolute age was instead derived by
\cite{vandenberg13} once again using the ACS data and theoretical isochrones; these authors
found a value of 12.75 Gyr, which, combined with its low metallicity, placed
NGC~6535 among the halo, accreted clusters in the age-metallicity plot.

\cite{testa01} presented the photometric data we used in our selection of
targets. They observed NGC~6535 with HST (Cycle 6, proposal 6625), using the
{\em Wide-Field Planetary Camera 2 (WFPC2)}, F555W, and F814W filters. These authors centred the cluster on the PC chip.
Since the small field of view of the WPFC2 did not cover the entire cluster,
they supplemented it with ground-based data (0.9m Dutch telescope, La Silla,
Chile, program 59.E-0532)  obtaining five $3.7\arcmin \times 3.7\arcmin$ fields
in the $V$ and $I$ filters.  

Using {\em ACS Treasury Program} data for GCs, \cite{milonehb} studied the HBs
and their relations with cluster parameters; referring to their Fig.~7, we would
expect an interquartile range (IQR) [O/Na] that is larger than about 0.5  dex(see
Sect.~\ref{nao}). NGC~6535 is also included in the HST {\em UV Legacy survey}
\citep{piottouv}; referring to their Fig.~16, the cluster displays a
normally split RGB (i.e. presence of MPs) and a general paucity of stars. This
is confirmed by \cite{milone17} who found a fraction of first-generation stars of
$54 \pm 8 \%$ using a total sample of 62 RGB stars.

\cite{leon00}  included NGC~6535 among the 20 Galactic GCs they investigated to
find tidal tails; as  for most of their sample, the cluster shows evidence of 
interactions and shocks with the Galactic plane/bulge in the form of tidal
extensions aligned with the tidal field gradient. \cite{halford_zaritsky} used
the HST data described above and measured an unusually flat (i.e. bottom-light)
present-day stellar mass function, which is at odds with the exceptionally high
mass-to-light ratio based on dynamical mass calculations. This could be due to
very strong external influence leading to mass stripping or to large amounts of
dark remnants; however, they were unable to clearly pinpoint the cause. Finally,
\cite{askar17} proposed that NGC~6535 has the kinematic and photometric
characteristic of what they call  a 'dark cluster', i.e. a cluster in which the
majority of the mass is presently locked in an intermediate black hole.

The spectroscopic material is much less abundant. \cite{zw84} used a
low-resolution, integrated spectrum to measure a radial velocity (RV) of
$-126\pm14$ km~s$^{-1}$ and determine [Fe/H]=$-1.75$ ($\sigma=0.15$). The same
technique was used by \cite{hsm86} to obtain RV=$-159\pm15$ km~s$^{-1}$. A very
different value is reported by \cite{pm93},  who give an average RV of
$-215.27\pm0.54$ km~s$^{-1}$, based on unpublished MMT and CTIO echelle spectra.
\cite{rutledge97a,rutledge97b} observed the near-IR calcium triplet region in
giant stars in 52 GCs (seven stars in NGC~6535); they found RV=$-204.8\pm14.0$ 
km~s$^{-1}$ and derived the cluster metallicity as [Fe/H]=$-1.78\pm0.07$ (on the
\citealt{zw84} scale) and $-1.51\pm0.10$ (on the \citealt{cg97} scale). Finally,
\cite{martell08}, in an effort to study carbon depletion in giants of 11 GCs,
took low-resolution spectra of two stars in NGC~6535  finding [C/Fe]=$-0.58$ and
$-0.29,$ respectively.

\begin{table}
\centering
\setlength{\tabcolsep}{1.mm}
\caption{Log of FLAMES observations.}
\begin{tabular}{lccccc}
\hline
Setup   &  UT Date   &  UT$_{init}$ & exptime & airmass & seeing\\
        & (yyyy-mm-dd) & (hh:mm:ss) & (s)          &        & (arcsec) \\
\hline
HR11  & 2014-07-01 & 07:03:25.899 & 3600  & 1.511 &  0.86\\
HR11  & 2014-07-02 & 03:30:04.291 & 3600  & 1.111 &  1.09\\
HR11  & 2014-07-25 & 04:27:17.459 & 3600  & 1.243 &  0.75\\
HR11  & 2014-07-29 & 02:10:50.669 & 3600  & 1.098 &  0.61\\
HR11  & 2014-08-01 & 04:05:18.439 & 3600  & 1.259 &  1.15\\
HR11  & 2014-08-02 & 00:47:02.373 & 3600  & 1.162 &  1.30\\ 

HR13  & 2014-08-02 & 03:00:05.225 & 3600  & 1.131 &  1.20\\
HR13  & 2014-08-02 & 04:41:08.280 & 3600  & 1.418 &  1.15\\
HR13  & 2014-08-03 & 01:11:16.019 & 3600  & 1.123 &  1.19\\
HR13  & 2014-08-03 & 02:25:19.678 & 3600  & 1.104 &  1.04\\
HR13  & 2014-08-03 & 03:38:27.712 & 3090  & 1.207 &  0.77\\
\hline
\end{tabular}
\label{log}
\end{table}

\section{Observations and analysis} \label{obs}

To select our targets for FLAMES we used the photometry by \cite{testa01},
downloading the catalogue from ViZier.  The $V,V-I$ CMD is shown in
Fig.~\ref{cmdoss}, lower panel. As expected from its Galactic position, the 
field stars contamination is conspicuous, but the cluster RGB and HB are
visible, especially when restricting to the very central region covered by HST.

We converted the positions, given as offsets with respect to the cluster centre,
to RA and Dec using stars in the  HST Guide Star Catalogue-II for the
astrometric conversion.\footnote{We used the code {\sc CataXcorr}, developed by
Paolo Montegriffo at the INAF - Osservatorio Astronomico di Bologna; see
http://www.bo.astro.it/$\sim$paolo/Main/CataPack.html} We then selected stars on
and near the RGB and allocated targets with the ESO positioner FPOSS. The
observed targets are indicated in Fig.~\ref{cmdoss}, upper panel. Given the
small field of view available and the positioner restrictions we were able to
observe only 45 stars (and about 30 sky positions).

\begin{figure}
\centering
\includegraphics[scale=0.43,bb=50 150 565 720,clip]{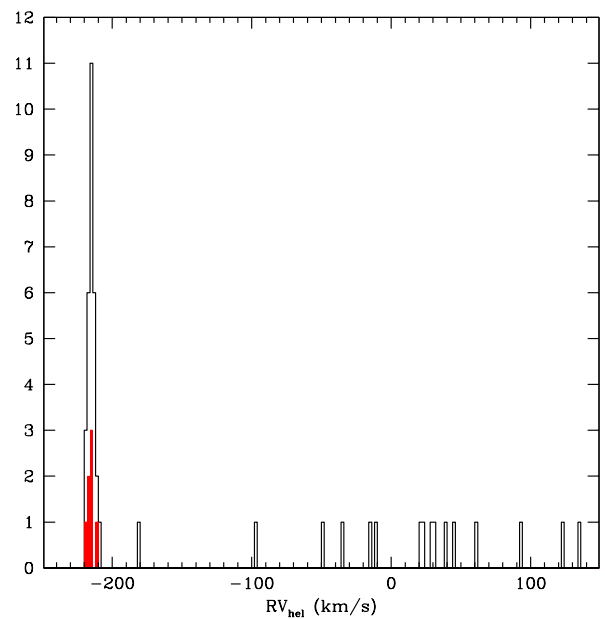}
\caption{Histogram of heliocentric RVs. The filled red histogram indicates the
seven UVES stars. The cluster stars are easily identified, with RV near $-215$
km~s$^{-1}$. }
\label{plotrv}
\end{figure}

\begin{table*}
\centering 
\setcounter{table}{1}
\caption{Information on the stars observed.}
\begin{tabular}{ccccccrcl}
\hline
\hline
 ID   &    RA       &  Dec         &  V     & V-I      &  K       &  RV    & err  & 2MASS ID \& Qflg\\
      & (hh:mm:ss)  & (dd:pp:ss)   &        &         & (2MASS)& (km~s$^{-1}$) & (km~s$^{-1}$)& \\
\hline
\multicolumn{9}{c}{UVES} \\
ESO1182 & 18:03:53.18 & -00:18:24.92  & 15.732 & 1.426 & 10.601  &   -211.66 &  0.19 & 18035317-0018247, AAA \\
ESO1217 & 18:03:52.22 & -00:18:12.90  & 16.046 & 1.378 & 10.856  &   -216.35 &  0.30 & 18035223-0018129, EEE \\
 ESO157 & 18:03:54.43 & -00:16:45.04  & 16.285 & 1.365 & 11.329  &   -219.44 &  0.52 & 18035444-0016449, AAA \\
ESO1601 & 18:03:51.08 & -00:14:39.89  & 16.097 & 1.417 & 11.706  &   -216.81 &  0.49 & 18035108-0014396, AAA \\
ESO2416 & 18:03:38.14 & -00:15:26.96  & 15.927 & 1.397 & 12.049  &   -215.86 &  0.31 & 18033813-0015267, AAA \\
   WF22 & 18:03:47.73 & -00:18:06.20  & 15.416 & 1.410 & 12.261  &   -215.96 &  0.23 & 18034771-0018059, AAA \\
   WF33 & 18:03:47.74 & -00:17:23.55  & 14.884 & 1.445 & 12.383  &   -215.10 &  0.19 & 18034773-0017234, AAA \\
\multicolumn{9}{c}{GIRAFFE, members} \\
ESO102  & 18:03:54.90 & -00:18:16.35  & 16.531 & 1.358 & 13.426  &   -213.30 &  0.35 & 18035489-0018161, AAA  \\
ESO1155 & 18:03:44.64 & -00:18:03.96  & 17.671 & 1.286 & 14.914  &   -214.11 &  0.55 & 18034463-0018037, AAB  \\
ESO1208 & 18:03:46.69 & -00:16:31.06  & 17.172 & 1.355 & 14.215  &   -213.75 &  0.34 & 18034668-0016308, AAA  \\
ESO2304 & 18:03:40.97 & -00:18:48.56  & 17.649 & 1.306 & 14.564  &   -209.99 &  0.42 & 18034096-0018483, AAA  \\
ESO2369 & 18:03:37.99 & -00:16:09.06  & 17.197 & 1.383 & 14.145  &   -213.34 &  0.84 & 18033799-0016087, AAA  \\
ESO24   & 18:03:44.10 & -00:18:31.27  & 17.453 & 1.303 & 14.590  &   -214.71 &  1.10 & 18034409-0018310, AAA  \\
ESO62   & 18:03:57.42 & -00:17:43.40  & 17.201 & 1.301 & 14.216  &   -214.96 &  1.32 & 18035741-0017431, AAA  \\
ESO659  & 18:03:43.48 & -00:20:34.96  & 16.359 & 1.399 & 13.084  &   -214.68 &  0.54 & 18034348-0020347, AAA  \\
ESO666  & 18:03:48.62 & -00:20:26.88  & 17.713 & 1.343 & 14.644  &   -216.48 &  0.39 & 18034862-0020265, AAA  \\
PC112   & 18:03:49.62 & -00:17:50.70  & 16.443 & 1.396 & 13.150  &   -217.15 &  0.51 & 18034961-0017505, AAA  \\
PC117   & 18:03:50.59 & -00:17:54.40  & 17.032 & 1.360 & 13.316  &   -213.56 &  0.51 & 18035058-0017538, AAA  \\
PC119   & 18:03:51.41 & -00:17:46.56  & 17.236 & 1.372 & 13.751  &   -214.86 &  0.45 & 18035139-0017467, AAA  \\
PC13    & 18:03:51.98 & -00:17:56.53  & 15.450 & 1.344 & 12.198  &   -214.52 &  1.09 & 18035199-0017564, AAA  \\
WF214   & 18:03:48.06 & -00:18:36.79  & 17.718 & 1.338 & 14.817  &   -211.00 &  0.83 & 18034805-0018366, AAB  \\
WF21    & 18:03:47.42 & -00:18:23.69  & 12.563 & 1.693 &  8.274  &   -216.59 &  0.72 & 18034741-0018234, AAA  \\
WF23    & 18:03:46.52 & -00:18:05.76  & 15.676 & 1.345 & 12.584  &   -217.25 &  4.25 & 18034651-0018054, AAA  \\
WF312   & 18:03:48.95 & -00:17:33.14  & 17.195 & 1.330 & 14.291  &   -218.28 &  0.86 & 18034895-0017331, AAA  \\
WF411   & 18:03:49.06 & -00:17:16.71  & 16.780 & 1.344 & 13.747  &   -215.08 &  0.84 & 18034904-0017160, AAA  \\
WF41    & 18:03:49.34 & -00:17:06.57  & 13.434 & 1.638 &  9.627  &   -215.64 &  0.66 & 18034932-0017059, AAA  \\
WF43    & 18:03:52.24 & -00:16:33.45  & 15.064 & 1.334 & 11.960  &   -212.09 &  0.90 & 18035223-0016330, AAA  \\
WF47    & 18:03:50.84 & -00:17:02.05  & 16.037 & 1.335 & 12.890  &   -213.21 &  0.87 & 18035083-0017015, AEA  \\
WF49    & 18:03:53.15 & -00:17:10.50  & 16.373 & 1.301 & 13.246  &   -218.34 &  1.33 & 18035315-0017101, AAA  \\
ESO803  & 18:03:45.38 & -00:20:13.23  & 17.020 & 1.306 & 13.966  &   -181.46 &  3.37 & 18034538-0020129, AUU, NM? \\
\multicolumn{9}{c}{GIRAFFE, non-members} \\
ESO121   &  18:03:54.65 & -00:17:02.78 & 17.343 & 1.464 &  &  38.92 &0.36 & \\
ESO1518  &  18:03:55.79 & -00:14:15.37 & 17.215 & 1.455 &  &  31.15 &0.51 & \\
ESO152   &  18:03:54.76 & -00:16:06.26 & 16.873 & 1.445 &  &  20.30 &1.63 & \\
ESO1588  &  18:03:55.50 & -00:15:02.35 & 15.301 & 1.444 &  & 134.75 &0.39 & \\
ESO2006  &  18:03:58.37 & -00:14:40.61 & 14.194 & 1.655 &  & -34.88 &0.07 & \\
ESO2048  &  18:04:00.96 & -00:18:33.59 & 16.265 & 1.321 &  &  23.66 &0.50 & \\
ESO2273  &  18:03:43.12 & -00:16:22.86 & 16.770 & 1.313 &  & 123.23 &0.47 & \\
ESO2320  &  18:03:40.35 & -00:18:11.16 & 16.631 & 1.455 &  & -48.77 &0.22 & \\
ESO2326  &  18:03:41.22 & -00:18:03.31 & 16.989 & 1.520 &  & -96.89 &0.50 & \\
ESO2334  &  18:03:28.75 & -00:17:44.19 & 17.022 & 1.431 &  & -10.14 &0.35 & \\
ESO395   &  18:03:49.84 & -00:15:41.79 & 17.183 & 1.440 &  &  44.42 &0.25 & \\
ESO647   &  18:03:52.86 & -00:20:45.65 & 16.184 & 1.274 &  & -15.00 &0.04 & \\
ESO819   &  18:03:44.05 & -00:19:55.54 & 17.216 & 1.449 &  &  28.91 &0.11 & \\
ESO99    &  18:03:58.38 & -00:18:31.82 & 15.463 & 1.563 &  &  60.59 &0.03 & \\
WF210    &  18:03:46.20 & -00:18:49.69 & 16.942 & 1.370 &  &  92.75 &0.77 & \\
\hline
\end{tabular}
\label{tabM}
\end{table*}

\subsection{FLAMES spectra}\label{spectra}
NGC~6535 was observed with the multi-object  spectrograph FLAMES@VLT
\citep{pasquini}. The observations were performed in service mode; a log is
presented in Table~\ref{log}.  We used the GIRAFFE high-resolution set-ups HR11
and HR13 (R=24200 and 22500, respectively), which contain two Na doublets and the
[O {\sc i}]  line at 6300 \AA, plus several Mg lines. The GIRAFFE observations
of 38 stars were coupled with the spectra of 7 stars obtained with the
high-resolution (R=47000) UVES 580nm set-up
($\lambda\lambda\simeq4800-6800$~\AA). Information on the 45 stars (ID,
coordinates, $V$, $V-I$, $K$, RV) is given in Table~\ref{tabM}.

The spectra were reduced using the ESO pipelines for UVES-FIBRE and GIRAFFE
data; they take care of bias and flat-field correction, order tracing,
extraction, fibre transmission, scattered light, and wavelength calibration. We
then used IRAF\footnote{IRAF is distributed by the National Optical Astronomical
Observatory, which are operated by the Association of Universities for Research
in Astronomy, under contract with the National Science Foundation.} routines on
the 1-D, wavelength-calibrated individual spectra to subtract the (average) sky,
correct for barycentric motion, combine all the exposures for each star, and
shift to zero RV.  The region near the [O {\sc i}] line was corrected for
telluric lines contamination before combining the exposures.  

The RV was measured via DOOp \citep{doop}, an automated wrapper for DAOSPEC
\citep{daospec} on the stacked spectra; the average heliocentric value for each
star is given in Table~\ref{tabM}, together with the rms. We show in
Fig.~\ref{plotrv} the histogram of the RVs; the cluster signature is evident
and we identified 29 of the 45 observed targets as cluster members on the basis
of their RV. Their average RV is $-214.97$ km~s$^{-1}$  (with $\sigma=2.22$), which is in
very good agreement with the value $-215.1\pm0.5$ reported by \cite{harris}.

\subsection{Atmospheric parameters}\label{paratmo} 
Only 30 stars (the 29 members plus one of more uncertain status) were retained
for further analysis.  We retrieved their 2MASS magnitudes \citep{2mass}, which
were used to determine the temperature, $K$ mag, 2MASS identification, and
quality flag are given in Table~\ref{tabM}.

Following our well-tested procedure \citep[for a lengthy description, see
e.g.][]{carretta09a,carretta09b}, effective temperatures $T_{\rm eff}$\ for our
targets were derived with an average relation between apparent $K$ magnitudes
and first-pass temperatures from $V-K$ colours and the calibrations of
\cite{alonso,alonso2}. This method permits us to decrease the star-to-star errors
in abundances due to uncertainties in temperatures, since magnitudes are less
affected by  uncertainties than colours.   The adopted reddening $E(B-V)=0.34$
and distance modulus $(m-M)_V=15.22$ are from the \cite{harris} catalogue; the
input metallicity [Fe/H]=$-1.79$ is from \cite{carretta09c} and \cite[][web
update]{harris}. Gravities were obtained from apparent magnitudes and distance
modulus, assuming the  bolometric corrections from \cite{alonso}. We adopted a
mass of 0.85~M$_\odot$\  for all stars and $M_{\rm bol,\odot} = 4.75$ as the
bolometric magnitude for the Sun, as in our previous studies.

We measured the equivalent widths (EW) of iron and other elements via the code
ROSA \citep{rosa}, adopting a relationship between EW and FWHM (for details, see
\citealt{bragaglia01}).  We eliminated trends in the relation between abundances
from Fe~{\sc i} lines and expected  line strength \citep{magain} to obtain
values of the microturbulent velocity $v_t$.  Finally, using the above values we
interpolated  within the \cite{kur} grid of model atmospheres (with the option
for overshooting on) to derive the final abundances, adopting for each star the
model with the appropriate atmospheric parameters and whose abundances matched
those derived from  Fe {\sc i} lines. The adopted atmospheric parameters (
$T_{\rm eff}$, $\log g$, [Fe/H], and $v_t$) are presented in
Table~\ref{tabatmo}.

\begin{figure*}
\centering
\includegraphics[scale=0.7]{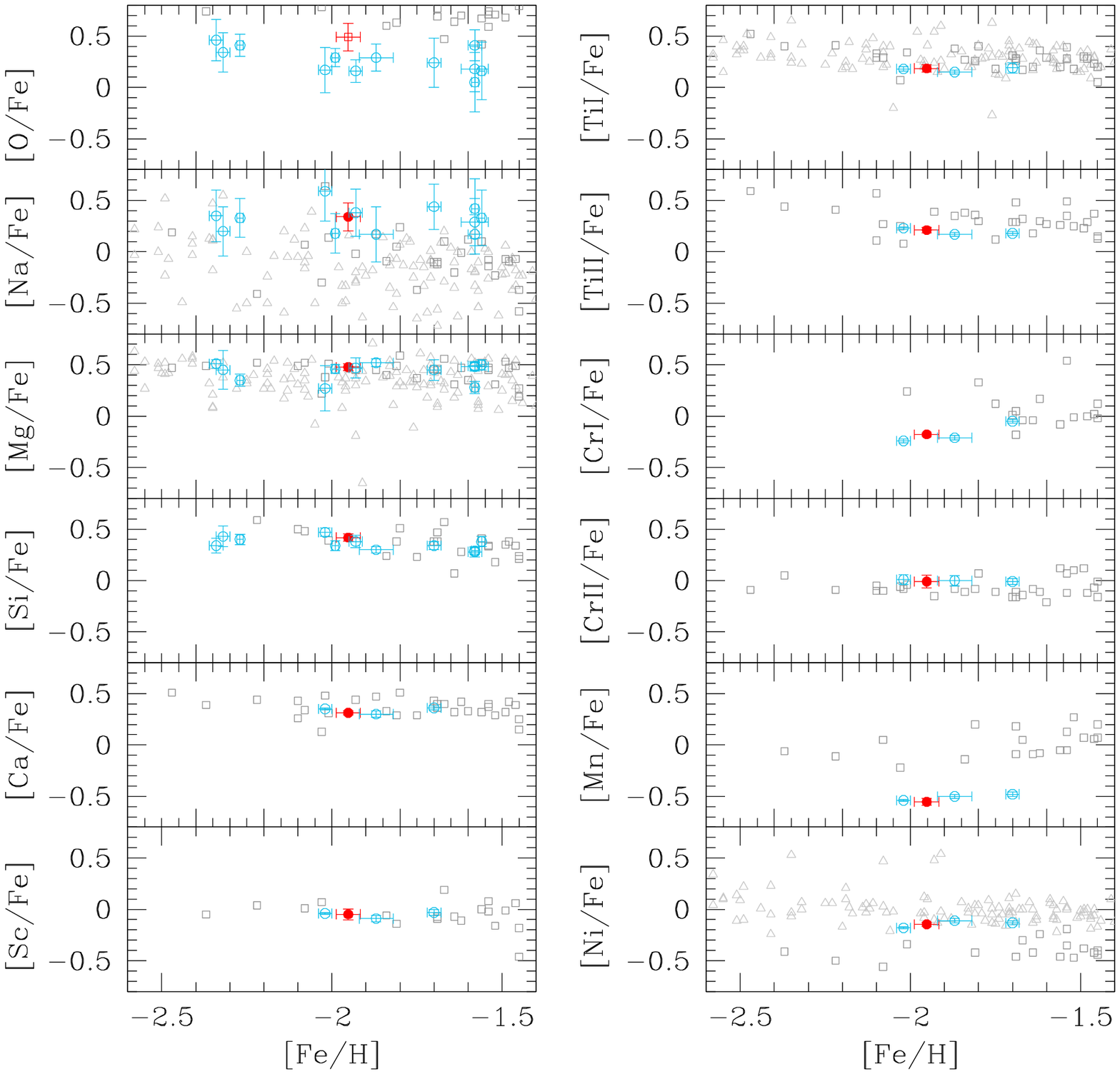}
\caption{Mean abundances for NGC~6535 (from Tab.~\ref{meanabu}) compared to
field stars  and other GCs of our FLAMES survey of similar metallicity. We use
here data from \cite{gratton03}, indicated by grey open squares, and
\cite{venn04}, indicated by light grey open triangles. For the GCs, indicated by
light blue open squares, abundances are taken from \cite{n4833} for NGC~4833,
\cite{m80} for NGC6093, \cite{n5634} for NGC~5634, and from \citet[][Tabs.~5 and
10, including only Fe, O, Na, Mg, and Si]{carretta09b} otherwise. The error bars
indicate the rms in each elemental ratio.}
\label{figfield}
\end{figure*}

\setcounter{table}{2}
\begin{table*}
\centering
\caption[]{Adopted atmospheric parameters and derived metallicity for confirmed member stars.}
\begin{tabular}{rccccrccrcc}
\hline
Star   &   $T_{\rm eff}$ &  $\log$ $g$ &  [A/H]  & $v_t$             &  nr &  [Fe/H]{\sc i} &  $rms$ &  nr &  [Fe/H{\sc ii} &  $rms$  \\
       &      (K)       &   (dex)     &  (dex)  & (km s$^{-1}$) &     &  (dex)     &       &     &  (dex)               \\
\hline
\multicolumn{11}{c}{UVES} \\
ESO1182  &4752 &2.19 &-1.96 &1.41 &37 &-1.958 &0.100 &10  &-1.904 &0.114  \\   
ESO1217  &4787 &2.27 &-1.92 &1.18 &37 &-1.919 &0.132 &10  &-1.895 &0.098  \\   
 ESO157  &4847 &2.43 &-2.00 &1.36 &26 &-1.997 &0.098 &10  &-1.975 &0.141  \\ \ 
ESO1601  &4830 &2.40 &-1.89 &1.37 &37 &-1.890 &0.115 & 8  &-1.843 &0.162  \\   
ESO2416  &4794 &2.31 &-1.97 &1.29 &34 &-1.969 &0.107 & 8  &-1.914 &0.053  \\   
   WF22  &4712 &2.10 &-1.97 &1.43 &40 &-1.973 &0.101 & 8  &-1.957 &0.040  \\   
   WF33  &4600 &1.80 &-1.96 &1.53 &56 &-1.959 &0.106 &13  &-1.958 &0.115  \\   
\multicolumn{11}{c}{GIRAFFE} \\
ESO102   &4901 &2.58 &-1.88 &0.80 &12 &-1.880 &0.153 & 2  &-1.917 &0.175  \\   
ESO1155  &5140 &3.16 &-1.92 &1.17 &11 &-1.914 &0.143 & 0  &99.999 &9.999  \\   
ESO1208  &5028 &2.92 &-1.95 &1.03 &15 &-1.948 &0.193 & 1  &-1.906 &9.999  \\   
ESO2304  &5084 &3.04 &-1.98 &0.27 &17 &-1.982 &0.189 & 0  &99.999 &9.999  \\   
ESO2369  &5017 &2.88 &-1.92 &1.03 &15 &-1.919 &0.141 & 1  &-1.945 &9.999  \\   
ESO24    &5088 &3.08 &-1.92 &1.03 & 9 &-1.915 &0.140 & 0  &99.999 &9.999  \\   
ESO62    &5028 &2.91 &-1.96 &0.93 &14 &-1.958 &0.169 & 0  &99.999 &9.999  \\   
ESO659   &4847 &2.42 &-2.04 &0.61 &23 &-2.038 &0.192 & 1  &-1.961 &9.999  \\   
ESO666   &5097 &3.07 &-1.91 &1.37 &15 &-1.907 &0.155 & 1  &-1.904 &9.999  \\   
PC112    &4857 &2.45 &-2.01 &1.29 &14 &-2.006 &0.177 & 1  &-2.010 &9.999  \\   
PC117    &4998 &2.84 &-1.99 &0.34 &10 &-1.992 &0.141 & 0  &99.999 &9.999  \\   
PC119    &5038 &2.94 &-2.01 &0.80 &16 &-2.007 &0.204 & 0  &99.999 &9.999  \\   
PC13     &4749 &2.11 &-2.06 &0.13 &18 &-2.063 &0.100 & 2  &-2.093 &0.059  \\   
WF214    &5125 &3.17 &-1.87 &0.73 &14 &-1.867 &0.172 & 0  &99.999 &9.999  \\   
WF21     &4075 &0.41 &-1.93 &2.17 &32 &-1.933 &0.150 & 3  &-1.884 &0.080  \\   
WF23     &4926 &2.29 &-1.97 &1.39 &21 &-1.970 &0.176 & 1  &-1.967 &9.999  \\   
WF312    &5040 &2.96 &-1.93 &0.95 &15 &-1.929 &0.123 & 1  &-1.915 &9.999  \\   
WF411    &4953 &2.72 &-2.07 &0.91 &13 &-2.066 &0.143 & 1  &-2.095 &9.999  \\   
WF41     &4292 &0.99 &-1.98 &1.56 &32 &-1.976 &0.140 & 1  &-1.955 &9.999  \\   
WF43     &4912 &2.04 &-1.98 &0.90 &26 &-1.977 &0.177 & 1  &-1.896 &9.999  \\   
WF47     &4864 &2.40 &-2.00 &1.17 &22 &-1.999 &0.155 & 0  &99.999 &9.999  \\   
WF49     &4886 &2.55 &-1.95 &0.83 &18 &-1.950 &0.146 & 1  &-1.900 &9.999  \\   
\hline
\end{tabular}
\label{tabatmo}
\end{table*}

\begin{table*}
\centering
\caption[]{Light element abundances for confirmed member stars. For O, lim=0
indicates an upper limit, lim=1 a measure. The PIE status is also indicated}
\begin{tabular}{rccccrcccrccc}
\hline
Star   &    nr &  [O/Fe] &  $rms$ &lim &  nr &  [Na/Fe] &  $rms$  &  PIE &  nr &  [Mg/Fe] &  $rms$\\
       & lines &(dex) &(dex)  &  & lines &(dex) &(dex) & & lines &(dex) &(dex)                 \\
\hline
\multicolumn{11}{c}{UVES} \\

ESO1182  &2  & 0.408 &0.086  &0   &2  & 0.177 &0.017  &P  &2  & 0.461 &0.091  \\ 
ESO1217  &1  & 0.588 & ---   &0   &2  & 0.426 &0.091  &I  &2  & 0.429 &0.093  \\ 
 ESO157  &1  & 0.699 & ---   &0   &2  & 0.323 &0.030  &I  &2  & 0.517 &0.230  \\ 
ESO1601  &1  & 0.496 & ---   &1   &2  & 0.403 &0.097  &I  &2  & 0.519 &0.161  \\ 
ESO2416  &1  & 0.401 & ---   &0   &2  & 0.461 &0.089  &I  &2  & 0.465 &0.192  \\ 
   WF22  &1  & 0.608 & ---   &0   &1  & 0.129 & ---   &P  &2  & 0.502 &0.120  \\ 
   WF33  &1  & 0.238 & ---   &1   &2  & 0.459 &0.004  &I  &2  & 0.451 &0.132  \\ 
\multicolumn{11}{c}{GIRAFFE} \\                                      
ESO102   &1  & 0.596 & ---   &0   &1  &-0.170 & ---   &P  &2  & 0.482 &0.320  \\ 
ESO1155  &0  & ---   & ---   &1   &-  & ---   & ---   &-  &1  & 0.436 & ---   \\ 
ESO1208  &1  & 0.600 & ---   &0   &2  & 0.485 &0.265  &I  &-  & ---   & ---   \\ 
ESO2304  &1  & 0.695 & ---   &0   &1  & 0.300 & ---   &I  &-  & ---   & ---   \\ 
ESO2369  &1  & 0.784 & ---   &0   &1  &-0.015 & ---   &P  &-  & ---   & ---   \\ 
ESO24    &1  & 0.852 & ---   &0   &1  & 0.299 & ---   &I  &1  & 0.411 & ---   \\ 
ESO62    &1  & 0.645 & ---   &1   &1  & 0.353 & ---   &I  &1  & 0.515 & ---   \\ 
ESO659   &1  & 0.472 & ---   &1   &2  & 0.549 &0.029  &I  &2  & 0.598 &0.236  \\ 
ESO666   &1  & 0.996 & ---   &0   &1  & 0.024 & ---   &P  &1  & 0.567 & ---   \\ 
PC112    &1  & 0.532 & ---   &1   &2  & 0.419 & ---   &I  &1  & 0.438 & ---   \\ 
PC117    &1  & 0.698 & ---   &1   &2  & 0.297 &0.104  &I  &1  & 0.510 & ---   \\ 
PC119    &1  & 0.918 & ---   &0   &0  & ---   & ---   &-  &-  & ---   & ---   \\ 
PC13     &1  & 0.414 & ---   &0   &2  & 0.374 &0.007  &I  &1  & 0.514 & ---   \\ 
WF214    &-  & ---   & ---   &1   &-  & ---   & ---   &-  &1  & 0.467 & ---   \\ 
WF21     &2  & 0.510 &0.002  &1   &2  & 0.344 &0.073  &I  &2  & 0.478 &0.091  \\ 
WF23     &1  & 0.846 & ---   &1   &2  &-0.017 &0.054  &P  &1  & 0.486 & ---   \\ 
WF312    &1  & 0.657 & ---   &0   &1  & 0.024 & ---   &P  &1  & 0.494 & ---   \\ 
WF411    &1  & 0.693 & ---   &1   &2  & 0.240 &0.016  &I  &-  & ---   & ---   \\ 
WF41     &1  & 0.210 & ---   &1   &4  & 1.032 &0.105  &I  &1  & 0.482 & ---   \\ 
WF43     &1  & 0.586 & ---   &1   &2  & 0.242 &0.035  &I  &1  & 0.342 & ---   \\ 
WF47     &1  & 0.407 & ---   &0   &3  & 0.618 &0.163  &I  &1  & 0.472 & ---   \\ 
WF49     &1  & 0.693 & ---   &1   &2  &-0.059 &0.076  &P  &1  & 0.470 & ---   \\ 
\hline
\end{tabular}
\label{tablight}
\end{table*}

\begin{table*}
\setlength{\tabcolsep}{1.mm}
\centering
\caption{Other elemental abundances for UVES spectra}
\begin{tabular}{rccc ccc ccc ccc ccc}
\hline
Star     &nr  &[Si/Fe] &rms &nr  &[Ca/Fe] &rms &nr &[Ti/Fe]I  &rms &nr &[Ti/Fe]II &rms &nr &[Sc/Fe]II &rms   \\
         & lines &(dex) &(dex) & lines &(dex) &(dex) & lines &(dex) &(dex) & lines &(dex) &(dex) & lines &(dex) &(dex) \\
\hline
ESO1182 &2 &0.408 &0.063 &15 &0.280 &0.090 & 7 &0.176 &0.131 &4 &0.164 &0.036 &4 &-0.113 &0.074   \\
ESO1217 &2 &0.455 &0.071 &15 &0.304 &0.144 & 8 &0.131 &0.159 &3 &0.236 &0.102 &2 &-0.094 &0.052   \\
 ESO157 &1 &0.465 &      &14 &0.305 &0.184 & 8 &0.195 &0.168 &3 &0.226 &0.077 &1 &-0.012 &        \\
ESO1601 &2 &0.379 &0.016 &15 &0.328 &0.105 & 8 &0.231 &0.135 &5 &0.202 &0.207 &2 &-0.101 &0.022   \\
ESO2416 &2 &0.418 &0.058 &15 &0.327 &0.155 & 8 &0.217 &0.130 &4 &0.187 &0.211 &2 & 0.003 &0.189   \\
   WF22 &  &      &      &15 &0.317 &0.097 & 8 &0.192 &0.120 &4 &0.233 &0.108 &3 & 0.012 &0.187   \\
   WF33 &2 &0.385 &0.073 &19 &0.328 &0.103 &14 &0.149 &0.094 &6 &0.229 &0.130 &8 &-0.029 &0.089   \\
\hline
Star      &nr &[Cr/Fe]I &rms &nr &[Cr/Fe]II &rms &nr  &[Mn/Fe]  &rms &nr &[Ni/Fe] &rms \\
         & lines &(dex) &(dex) & lines &(dex) &(dex) & lines &(dex) &(dex) & lines &(dex) &(dex) & & & \\
\hline
ESO1182  &6 &-0.177 &0.086 &2 &-0.010 &0.029 &1 &-0.589 &      &5 &-0.127 &0.105 &&& \\
ESO1217  &6 &-0.197 &0.119 &2 &-0.035 &0.037 &1 &-0.588 &      &3 &-0.126 &0.120 &&&\\
 ESO157  &6 &-0.178 &0.112 &1 & 0.044 &      &1 &-0.546 &      &3 &-0.162 &0.100 &&&\\
ESO1601  &7 &-0.168 &0.122 &2 &-0.128 &0.079 &1 &-0.568 &      &4 &-0.141 &0.115 &&&\\
ESO2416  &7 &-0.158 &0.135 &2 & 0.038 &0.008 &1 &-0.569 &      &3 &-0.172 &0.013 &&&\\
   WF22  &6 &-0.164 &0.157 &2 &-0.019 &0.079 &1 &-0.495 &      &5 &-0.138 &0.180 &&&\\
   WF33  &7 &-0.194 &0.106 &3 & 0.038 &0.083 &3 &-0.524 &0.224 &9 &-0.156 &0.092 &&&\\
\hline
\end{tabular}
\label{tabuves}
\end{table*}

\section{Abundances}\label{abu}
In addition to iron, we present here results for the light elements O, Na, and Mg for the
entire sample of stars.  The abundance ratios for these elements are given in
Table~\ref{tablight}, together with number of lines used and rms scatter.   

For the seven stars observed with UVES, we also provide abundances for Si, Ca,
Ti {\sc i,ii}, Sc {\sc ii}, Cr {\sc i,ii}, Mn, and Ni; these abundances are presented in
Table~\ref{tabuves}. Since the cluster presents some peculiarities, it is important to see whether these also extend
to the whole pattern of abundances. Some of the peculiarities of the cluster include that it is
metal poor but located in the inner Galaxy, it is perhaps the lowest mass GC to
show MPs, and the Na-O anti-correlation seems more extended than expected from its
present-day mass; see next Section.  The cluster mean elemental ratios are
plotted in Fig.~\ref{figfield} in comparison to field stars and other GCs of
our FLAMES survey of similar metallicity. We do not see any anomalous
behaviour; NGC~6535 seems to be a normal (inner) halo GC; see also below. Hence, in
the present paper we dwell only on the light element abundances and we defer
any further discussion on heavier species and all neutron-capture elements to
future papers.

The abundances were derived  using EWs. The atomic data for the lines  and solar reference values  come from \cite{gratton03}.  The Na abundances were
corrected for departure from local thermodynamical equilibrium according to
\cite{gratton99}, as in all the other papers of our FLAMES survey.

To estimate the error budget we closely followed the procedure described in
\cite{carretta09a,carretta09b}. Table~\ref{tabsensu} (for UVES spectra) and
Table~\ref{tabsensg} (for GIRAFFE spectra) provide the sensitivities of
abundance ratios to errors in atmospheric parameters and EWs and the internal
and systematic errors.   For systematic errors we mean  the errors that are
different for the various GCs considered in our series and that produce
scatter in relations involving different  GCs; however, they do not affect  the
star-to-star scatter in any given GC.  The sensitivities were obtained by
repeating the abundance analysis for all stars,  while changing one atmospheric
parameter at the time, then taking the average; this was carried out separately for
UVES and GIRAFFE spectra. The amount of change in the input parameters used in
the sensitivity computations is given in the Table header. 

The derived Fe abundances do not show any trend with temperature and gravity
and the neutral and ionized species give essentially the same value, as do
results based on GIRAFFE or UVES spectra (see Table~\ref{meanabu}). In fact, we
obtain the following mean values: 
[Fe/H]{\sc i}=-1.952$\pm0.006$ (rms=0.036), 
[Fe/H]{\sc ii}=-1.921$\pm0.008$ (rms=0.046) dex for the 7 UVES stars; and 
[Fe/H]{\sc i}=-1.963$\pm0.003$ (rms=0.053), 
[Fe/H]{\sc ii}=-1.953$\pm0.005$ (rms=0.069) dex for the 22 GIRAFFE stars. 

This the first high-resolution spectroscopic study of this cluster, and hence no
real comparison with previous determinations is possible. However, the
metallicity we find is in reasonable agreement with that based on CaT
\citep[$-1.5$ or $-1.8$][]{rutledge97b}. The metallicity is also consistent
with  the value in \citet[][i.e. -1.79]{harris}, which comes from
\cite{carretta09c}. In that paper we built a metallicity scale based on GCs
observed at high resolution (NGC~6535 was not among these GCs) and recalibrated
other scales. 

Several studies have shown that the age-metallicity relation of MW GCs is
bifurcated with one sequence of old, essentially coeval GCs at all
metallicities and a second sequence of metal-richer GCs; see e.g.
\cite{vandenberg13,leaman13} and references therein. These sequences are
populated by clusters associated with the halo and the disk, respectively (see
Fig.~2 in \citealt{leaman13}, in which Rup~106, Ter~7, Pal~12, and
NGC~6791 are highlighted, which are discussed in Sect.~\ref{massage}). The
old age and the confirmed low metallicity, place NGC~6535 in the sequence of
halo, accreted GCs; see e.g. \cite{vandenberg13,leaman13}. 

\begin{table*}
\centering
\caption[]{Sensitivities of abundance ratios to variations in the atmospheric
parameters and to errors in the equivalent widths and errors in abundances for
stars of NGC~6535 observed with UVES.}
\begin{tabular}{lrrrrrrrr}
\hline
Element     & Average  & T$_{\rm eff}$ & $\log g$ & [A/H]   & $v_t$    & EWs     & Total   & Total      \\
            & n. lines &      (K)      &  (dex)   & (dex)   &kms$^{-1}$& (dex)   &Internal & Systematic \\
\hline        
Variation&             &  50           &   0.20   &  0.10   &  0.10    &         &         &            \\
Internal &             &   3           &   0.04   &  0.04   &  0.10    & 0.02    &         &            \\
Systematic&            &  61           &   0.06   &  0.09   &  0.04    &         &         &            \\
\hline
$[$Fe/H$]${\sc  i}& 38 &    +0.074  & $-$0.019 &$-$0.014 & $-$0.023 & 0.018  &0.030 &0.092 \\
$[$Fe/H$]${\sc ii}& 10 &  $-$0.008  &   +0.075 &  +0.017 & $-$0.016 & 0.034  &0.041 &0.031 \\
$[$O/Fe$]${\sc  i}&  1 &  $-$0.048  &   +0.094 &  +0.038 &   +0.022 & 0.108  &0.113 &0.088 \\
$[$Na/Fe$]${\sc i}&  2 &  $-$0.042  & $-$0.014 &  +0.002 &   +0.019 & 0.076  &0.079 &0.073 \\
$[$Mg/Fe$]${\sc i}&  2 &  $-$0.029  & $-$0.016 &  +0.002 &   +0.006 & 0.076  &0.077 &0.038 \\
$[$Si/Fe$]${\sc i}&  2 &  $-$0.045  &   +0.029 &  +0.009 &   +0.016 & 0.076  &0.078 &0.058 \\
$[$Ca/Fe$]${\sc i}& 15 &  $-$0.025  & $-$0.001 &  +0.001 &   +0.003 & 0.028  &0.028 &0.031 \\
$[$Sc/Fe$]${\sc ii}& 3 &    +0.019  & $-$0.002 &  +0.002 &   +0.006 & 0.062  &0.063 &0.031 \\
$[$Ti/Fe$]${\sc i}&  9 &    +0.012  & $-$0.003 &$-$0.004 & $-$0.005 & 0.036  &0.036 &0.020 \\
$[$Ti/Fe$]${\sc ii}& 4 &    +0.017  & $-$0.003 &  +0.000 &   +0.002 & 0.054  &0.054 &0.023 \\
$[$V/Fe$]${\sc i} &  1 &    +0.014  &   +0.001 &$-$0.001 &   +0.021 & 0.108  &0.110 &0.095 \\
$[$Cr/Fe$]${\sc i}&  6 &    +0.003  &   +0.000 &$-$0.003 & $-$0.005 & 0.044  &0.044 &0.007 \\
$[$Cr/Fe$]${\sc ii}& 2 &  $-$0.005  & $-$0.002 &$-$0.007 &   +0.010 & 0.076  &0.077 &0.024 \\
$[$Mn/Fe$]${\sc i}&  1 &  $-$0.011  &   +0.001 &  +0.000 &   +0.004 & 0.108  &0.108 &0.019 \\
$[$Ni/Fe$]${\sc i}&  5 &  $-$0.012  &   +0.017 &  +0.005 &   +0.012 & 0.048  &0.050 &0.018 \\
\hline
\end{tabular}
\label{tabsensu}
\end{table*}

\begin{table*}
\centering
\caption[]{Sensitivities of abundance ratios to variations in the atmospheric
parameters and errors in the equivalent widths, and errors in abundances for
stars of NGC~6535 observed with GIRAFFE.}
\begin{tabular}{lrrrrrrrr}
\hline
\hline
Element     & Average  & T$_{\rm eff}$ & $\log g$ & [A/H]   & $v_t$    & EWs     & Total   & Total      \\
            & n. lines &      (K)      &  (dex)   & (dex)   &kms$^{-1}$& (dex)   &Internal & Systematic \\
\hline        
Variation&             &  50           &   0.20   &  0.10   &  0.10    &         &         &            \\
Internal &             &   3           &   0.04   &  0.05   &  0.25    & 0.04    &         &            \\
Systematic&            &  62           &   0.06   &  0.08   &  0.05    &         &         &            \\
\hline
$[$Fe/H$]${\sc  i} & 17  &   +0.059 & $-$0.010 & $-$0.010 & $-$0.018 & 0.038 & 0.059 & 0.075 \\
$[$Fe/H$]${\sc ii} &  1  & $-$0.015 &   +0.078 &   +0.012 & $-$0.003 & 0.158 & 0.159 & 0.035 \\
$[$O/Fe$]${\sc  i} &  1  & $-$0.032 &   +0.085 &   +0.028 &   +0.018 & 0.158 & 0.166 & 0.063 \\
$[$Na/Fe$]${\sc i} &  2  & $-$0.030 & $-$0.013 &   +0.000 &   +0.016 & 0.112 & 0.119 & 0.082 \\
$[$Mg/Fe$]${\sc i} &  1  & $-$0.028 &   +0.004 &   +0.001 &   +0.013 & 0.158 & 0.161 & 0.038 \\
\hline
\end{tabular}
\label{tabsensg}
\end{table*}

\begin{figure}
\centering
\includegraphics[scale=0.44]{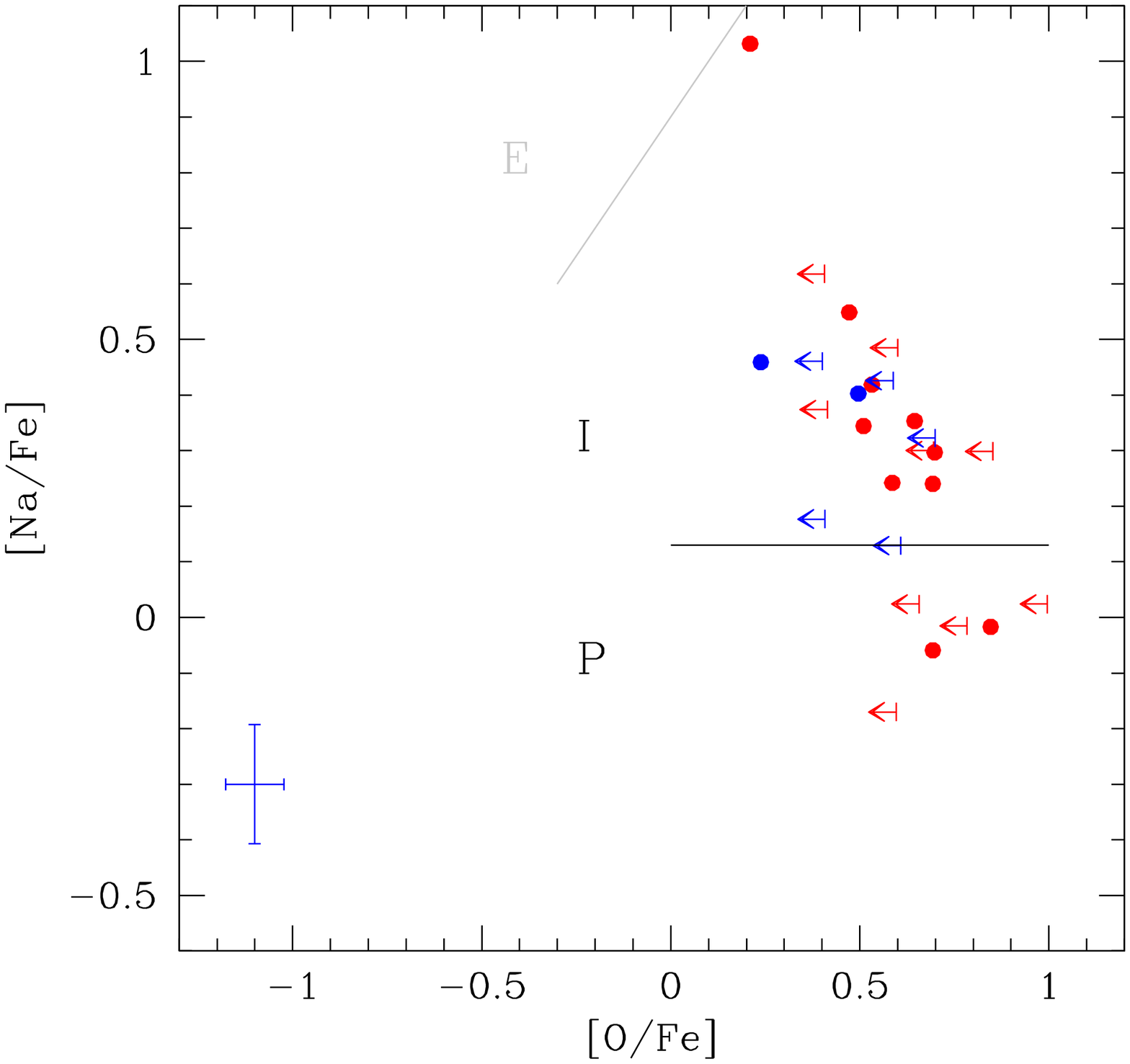}
\caption{Abundances of Na and O for NGC~6535; red symbols indicate GIRAFFE
spectra and blue symbols UVES spectra. Upper limits in O are also indicated.
We also show the separation in PIE components; no E star is present
in this cluster.}
\label{fignao}
\end{figure}

\section{ Na-O anti-correlation}\label{nao}

Unfortunately, we could only obtain upper limits in O abundance for many stars,
but this did not compromise the main goal of our work, i.e. detecting (or not)
MPs in this low-mass cluster. Indeed, NGC~6533 turned out to be a normal MW GC,
showing the usual anti-correlation between O and Na; see Fig.~\ref{fignao}.  This would not have been evident using only the seven
stars observed with UVES. We need to be cautious in relying on small number
statistics to  exclude the presence of MPs (e.g. Terzan~7, Pal~12).

\begin{figure}
\centering
\includegraphics[scale=0.6]{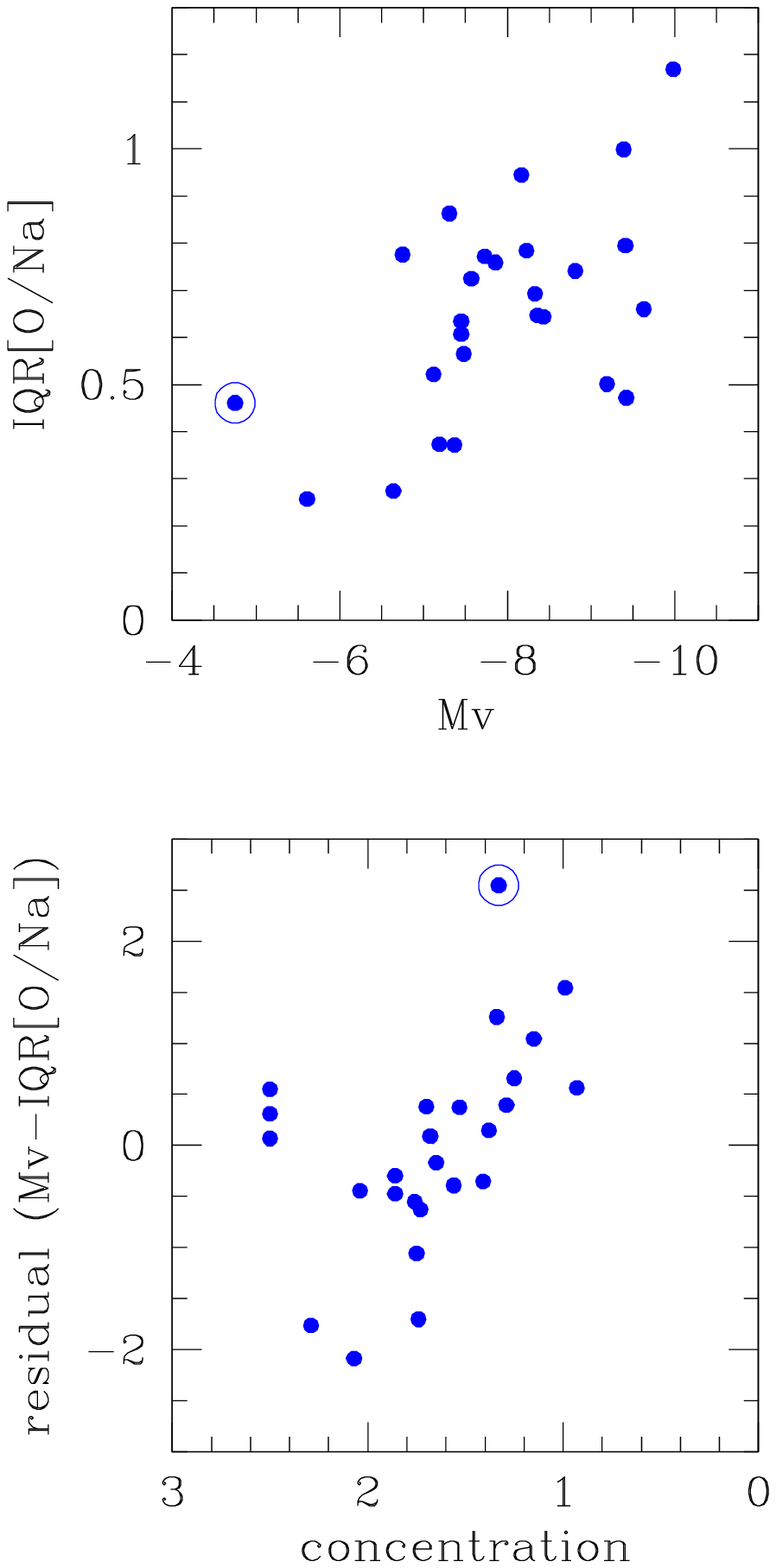}
\caption{Upper panel: IQR[O/Na] as a function of M$_V$ for all the GCs in our
FLAMES survey. Lower panel: Residuals around the relation between M$_V$ and
IQR[O/Na] as a function of the cluster concentration $c$ \citep{harris}. The three GCs at $c=2.5$ are post-core collapse systems. In both
panels NGC~6535 is indicated by a large open circle.}
\label{figiqr}
\end{figure}

The extension of the Na-O anti-correlation in NGC~6535 can be measured with
the interquartile ratio of [O/Na] \citep[see][]{carretta06}; from the 26 stars
with both Na and O abundances, we find IQR[O/Na]=0.44.  This value  is close to
the expectation based on HST photometry of its horizontal branch \citep[][see
Sect.~\ref{lit}]{milonehb}, but looks large for the present cluster mass,
according to the relation between IQR[O/Na] and absolute $V$ magnitude M$_V$ we
found (see Fig.~\ref{figiqr}, upper panel). However, this is true also in
other cases in which a strong mass loss is suspected, such as NGC~288, M71, and
NGC~6218. In \cite{n4833} we suggested that cluster concentration ($c$)
also plays a role and may explain part of the scatter in the IQR-M$_V$ relation. If
we plot the residuals around this relation against $c$, we see a neat
anti-correlation (Fig.~\ref{figiqr}, lower panel). The outliers are three
post-core collapse GCs, for which the concentration parameter is
arbitrarily set at 2.5 since they do not fit a King luminosity profile, and
NGC~6535, which again indicates some peculiarity for this cluster.

Using the separation defined in \cite{carretta09a} of FG and SG stars in  the P, I,
and E populations (i.e.  primordial, intermediate, and extreme, respectively), we
see that NGC~6535 does not harbour any E star but has a large number of I
stars. The lack of E stars is consistent with the low mass of the cluster  and also suggests
that its initial mass should not have been too large. The fraction of P
(FG) and I (SG) stars turns out $31\pm10\%$ and $69\pm16\%$, respectively, which is in
line with the other GCs of our survey \citep{zcorri}. 

From our extensive FLAMES survey, we derived an almost constant
\citep[e.g.][and following works]{carretta09a,zcorri} fraction of FG to SG
stars with FG stars constituting about one-third of the present-day cluster
population, according to Na and O abundances (confirmed also by
\citealt{bastian_lardo}). This is valid at least for the more massive GCs
(Terzan~8 is an exception, see \citealt{ter8}), even with some variation, from
about 25\% to 50\%. Also, a different fraction could be obtained
with a  separation based on photometry; see the case of NGC~288 discussed in 
\cite{stromgren} or M~13 \citep{massari16}. \cite{milone17} used the HST UV
Legacy survey to characterize MPs using what they called a ``chromosome map". They
were able to measure the fraction of FG stars (indicated as 1G) and found it
variable from cluster to cluster,   from $\sim 8\%$ to $\sim 67\%$. However,
their median value is 36\% (rms=0.09) and if we compare their values and what
we obtain from our Na-O sample for the GCs in common, 1G and P fractions
generally follow a one-to-one relation (Carretta et al., in preparation). For
NGC~6535, their $54\pm8$ is only marginally inconsistent with our $31\pm10\%$
value for 1G/P stars.  According to \cite{milone17} there is a correlation of
the MP phenomenon with cluster mass with higher mass GCs showing a lower
frequency of 1G stars. While this  closely matches what we found for the
extension of the Na-O anti-correlation, larger in more massive GCs
(\citealt{zcorri}, see also Fig.~\ref{figiqr}), we do not find a relation
between the fraction of FG stars and cluster mass (or age), when the separation
in populations is based on Na and O.

\begin{figure}
\centering
\includegraphics[scale=0.44]{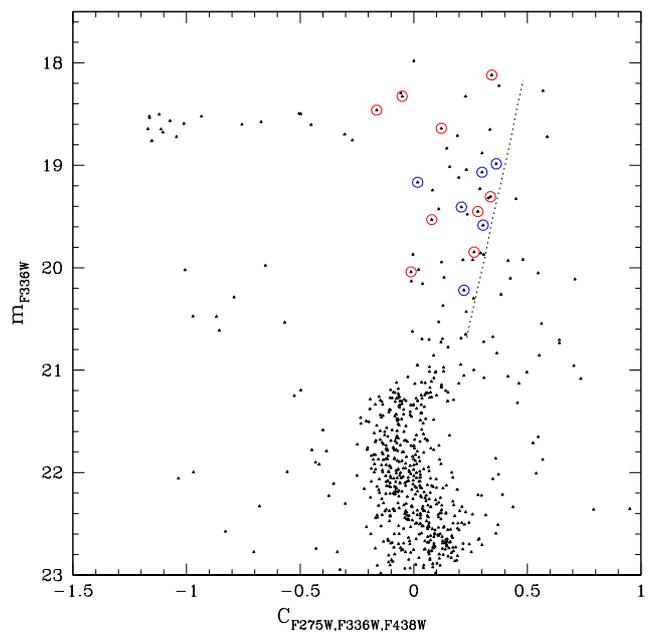}
\caption{Plot of the HST UV Legacy Survey data for NGC~6535, using the
pseudo-colour $C_{F275W,F336W,F438W}$ against the $F336W$ mag. First (P) and
second generation (I) stars in common with our spectroscopic sample are
indicated by large blue and red circles, respectively. They tend to segregate in
colour along two sequences.}
\label{figuv}
\end{figure}

\begin{figure}
\centering
\includegraphics[bb=20 410 580 700,clip,width=8cm]{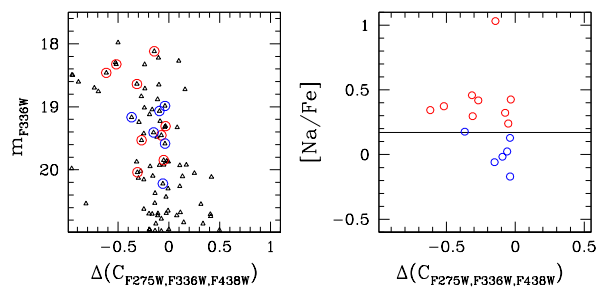}
\caption{Upper panel: $\Delta {\rm C_{F275W,F336W,F438W}}$ obtained using the
line in the previous figure vs. F336W. Lower panel: The same $\Delta{\rm
C_{F275W,F336W,F438W}}$ colour, plotted against [Na/Fe]. The horizontal lines
indicates the separation between P and I stars (see Fig.~\ref{fignao}).}
\label{figuv2}
\end{figure}

Also because photometry and high-resolution spectroscopy do not seem to tell
exactly the same story, it is interesting to compare their  results for
NGC~6535. We downloaded the HST UV Legacy early-release data from the Mikulski
Archive for Space Telescopes (MAST) and cross-matched the catalogue with our
stars, finding only 15 objects in common. Figure~\ref{figuv} shows a diagram
plotting the pseudo-colour $C_{F275W,F336W,F438W} = F275W-2\times F336W+F438W$ 
against F336W magnitude with our stars indicated by large open circles.  We
did not apply any quality selection to the photometry or try differential
reddening correction, and therefore the  RGB separation in two branches is less evident
than in \cite{piottouv}.  We may see that P and I separate  along different
sequences, although not perfectly. To better appreciate this, we used a line to
straighten the RGB and compute the difference in pseudo-colour, i.e.
$\Delta(C_{F275W,F336W,F438W})$. The left-hand panel of Fig.~\ref{figuv2} shows
the result; all P stars (with one exception) are confined to the right side and
I stars tend to occupy the left side of the RGB, but are more spread out. If we
consider Na abundance (right-hand panel of Fig.~\ref{figuv2}), we see that five
of the P stars share the same value (mean [Na/Fe]=$-0.02$, rms=0.11). The
exception is exactly at the border of our P, I separation and may also be a
misclassification. Instead, at about the same sodium, I stars may have a wide
range of $\Delta(C_{F275W,F336W,F438W})$ (that is, a range in C, N abundance),
another manifestation that Na and N are correlated, but do not tell exactly the
same story, as previously stated, e.g. by \cite{smith13,smith15}.

\begin{table}
\centering
\caption{Mean abundances in NGC~6535.}
\setlength{\tabcolsep}{1.5mm}
\begin{tabular}{llllrcc}
\hline
Element              &  stars & mean &rms & stars & mean &rms\\
    & \multicolumn{3}{c}{UVES}  & \multicolumn{3}{c}{GIRAFFE} \\
\hline
$[$O/Fe$]${\sc i}    & 7& +0.491 & 0.134 &20& +0.639 & 0.184     \\
$[$Na/Fe$]${\sc i}   & 7& +0.340 & 0.136 &20& +0.314 & 0.320 \\
$[$Mg/Fe$]${\sc i}   & 7& +0.478 & 0.035 &17& +0.470 & 0.057  \\
$[$Si/Fe$]${\sc i}   & 6& +0.418 & 0.035 &&&  \\
$[$Ca/Fe$]${\sc i}   & 7& +0.312 & 0.018 &&&  \\
$[$Sc/Fe$]${\sc ii}  & 7& -0.048 & 0.053 &&&  \\
$[$Ti/Fe$]${\sc i}   & 7& +0.184 & 0.036 &&&  \\
$[$Ti/Fe$]${\sc ii}  & 7& +0.211 & 0.027 &&&  \\
$[$Cr/Fe$]${\sc i}   & 7& -0.177 & 0.015 &&&  \\
$[$Cr/Fe$]${\sc ii}  & 7& -0.010 & 0.061 &&&  \\
$[$Mn/Fe$]${\sc i}   & 7& -0.554 & 0.035 &&&  \\
$[$Fe/H$]${\sc i}    & 7&-1.952  & 0.036 &22 &-1.963 & 0.054 \\
$[$Fe/H$]${\sc ii}   & 7&-1.921  & 0.046 &14 &-1.953 & 0.069 \\
$[$Ni/Fe$]${\sc i}   & 7&-0.146  & 0.018 &&&  \\
\hline
\end{tabular}
\label{meanabu}
\tablefoot{Solar reference values are
Fe {\sc i} 7.54, Fe {\sc ii} 7.49, O 8.79, Na 6.21, Mg 7.43, Si 7.53, Ca 6.27, Sc 3.13, Ti {\sc i} 5.0, Ti {\sc ii} 5.07, Cr {\sc i} 5.67, Cr {\sc ii 5.71}, Mn 5.34, Ni 6.28, from \cite{gratton03}. }
\end{table}

\begin{figure}
\centering
\includegraphics[scale=0.44]{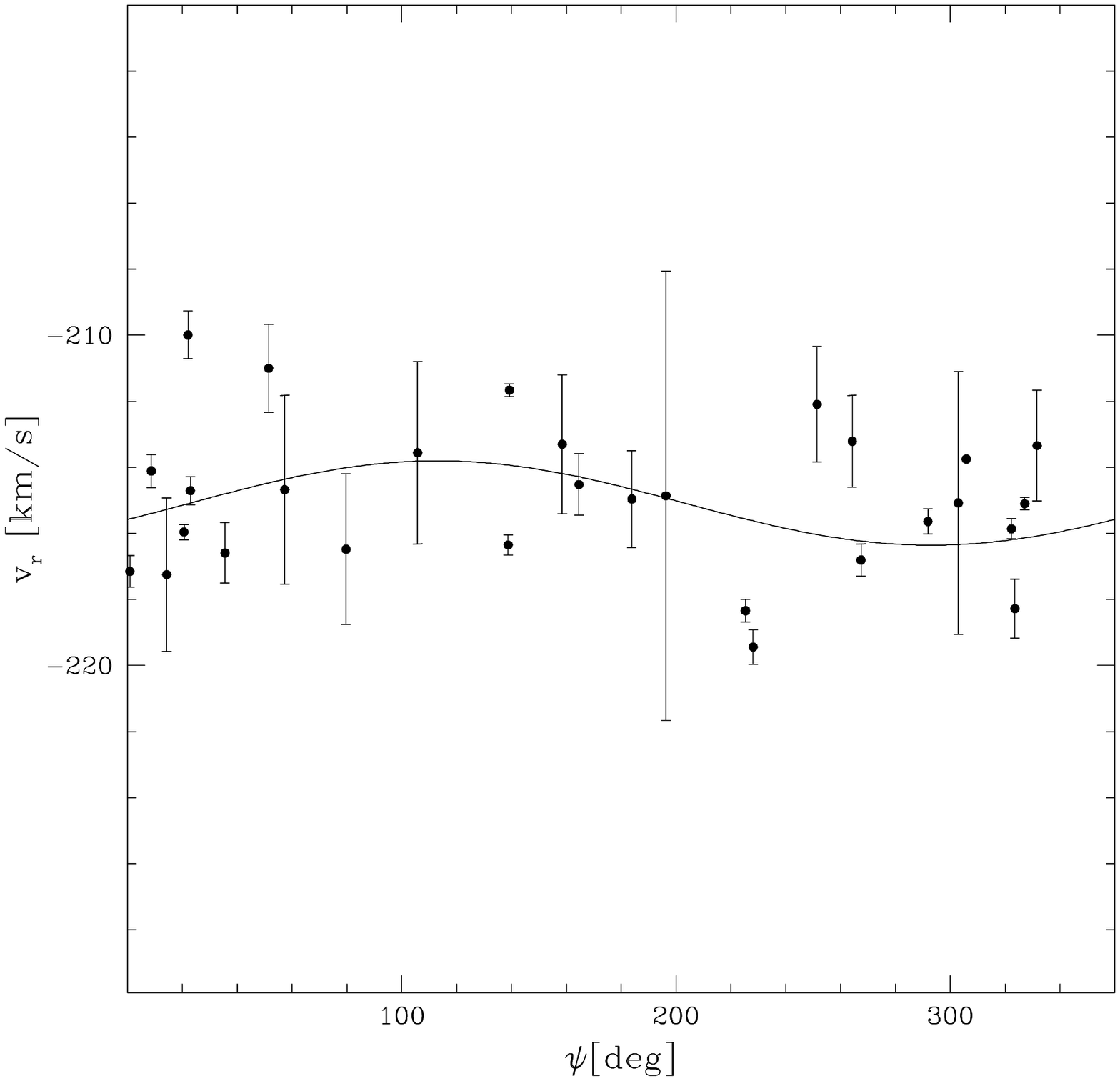}
\caption{Systemic rotation of NGC~6535. The best-fit  solution is compatible
with negligible rotation.}
\label{figrot}
\end{figure}

\section{Internal kinematics}\label{kin}

In spite of the relatively small number of cluster members the present dataset 
represents the most extensive set of radial velocities for NGC~6535 and is very
useful to study the internal kinematics of this cluster.

As a first step, we test the presence of systemic rotation. For this purpose, 
in Fig.~\ref{figrot} the radial velocities of the 29 {\it bona fide} members
are plotted against their position angles. The best-fit sinusoidal curve
indicates a rotation amplitude of $A_{rot}~sin i=1.28 \pm 2.71$ km~s$^{-1}$,
compatible with no significant rotation.

We then used our radial velocity dataset to estimate the dynamical mass of the
system. For this purpose we fitted the distribution of radial velocities with
both a single mass \cite{king66} model and a multi-mass King-Michie models
\citep{gunn_griffin}. In particular, for each model we tuned the model mass to
maximize the log-likelihood  $$L=-\sum_{i=1}^{N}\left(\frac{(v_{i}-\langle v
\rangle)^{2}}{(\sigma_{i}^{2}+\epsilon_{i}^{2})}+ln
(\sigma_{i}^{2}+\epsilon_{i}^{2})\right),$$ where N(=29) is the number of
available radial velocities, $v_{i}$ and $\epsilon_{i}$ are  the radial
velocity of the i-th star and its associated uncertainty, and $\sigma_{i}$ is
the line-of-sight velocity dispersion predicted by the model at the distance
from the cluster centre of the i-th star.  The best-fit single mass model
provides a total mass of $2.12\times10^{4} M_{\odot}$. For the multi-mass model
we chose a present-day mass function of single stars with a positive slope of
$\alpha=+1$ (for reference, a Salpeter mass function has $\alpha=-2.35$). Such
a peculiar mass function is indeed necessary to reproduce the paucity of
low-mass stars observed in the HST CMD presented by \cite{halford_zaritsky}. We
adopted the prescriptions for dark remnants and binaries of Sollima et al.
(2012) assuming a binary fraction of 4\% and a flat distribution of
mass ratios. The derived mass turned out to be 2.21$\times10^{4} M_{\odot}$ with
a typical uncertainty of $\sim 7.8\times10^{3} M_{\odot}$. This value is larger
than found by \cite{mlvdm05}, who have Log(mass)=3.53; the difference comes
from the different methods used (kinematics versus photometric profiles). In
any case, even our value places NGC~6535 among the presently less massive MW GCs.

By assuming the absolute magnitude $M_{V}=-4.75$ listed for this cluster in the
Harris catalogue  (\citealt{harris}, 2010 edition), the corresponding $M/L_{V}$
ratio are 3.12 and 3.25 for the single mass and multi-mass models,
respectively, while  larger M/L ratios (3.78 and 3.95) are  instead obtained if
the integrated magnitude $M_{V}=-4.54$ by \cite{mlvdm05} is adopted. Such a
large M/L ratio is consistent with what was already measured in other low-mass
clusters of our survey \cite[e.g.][]{ter8}.  This evidence can be
interpreted as an effect of the strong interaction of this cluster with the
Galactic tidal field affecting the M/L in a twofold way: {\it i)} the tidal
heating inflates the cluster velocity dispersion spuriously increasing the
derived dynamical mass, and {\it ii)} the efficient loss of stars leads to an
increased relative fraction of remnants contributing to the mass without
emitting any light.  Dedicated N-body simulations are needed to understand the
relative impact of the two above-mentioned effects.

\section{Age and mass limits for multiple populations}\label{massage}

As already put forward in the Introduction, there seems to be an observational
lower limit to the mass of a cluster to display the Na-O anti-correlation
\citep{zcorri}. We used the absolute magnitude in the $V$ band, M$_V$, as a
proxy for mass, since it is available for all MW GCs \cite[see the  catalogue
by][]{harris} while the mass is not. However, the majority
of clusters for which information on the presence (or, very rarely, absence) of
anti-correlations between light elements is available have high mass. We then
decided to observe systematically lower mass GCs along with old and massive OCs
to obtain and analyse homogeneously samples of stars as large as those for the other
clusters of our survey. We obtained data on a few objects (see Introduction),
but also other groups were active on the same line and therefore the situation improved
in recent years.

We revise here the census of clusters for which indication on MPs has been
obtained. Starting from the clusters in \cite{zcorri} (see their Fig.~3) and
the updated table in \cite{krause16}, we scanned the literature up to June 2017
for new results on the anti-correlation of light elements in GCs based on
high-resolution spectroscopy (i.e. involving Na, O or Mg, Al). At variance
with \cite{zcorri}, we also considered evidence of MPs based on photometry or
low-resolution spectroscopy (i.e. involving C, N), whenever no high-resolution
spectroscopy was available for the particular cluster. For the OCs, we
only considered a subsample: the two old, massive OCs Berkeley~39 and
NGC~6791\footnote{For this cluster, \cite{geisler12} proposed the presence of
MPs, which we and \cite{cunha15} did not confirm. We consider it to be a single
stellar population unless otherwise proved by new observations.} observed and
analyzed homogeneously by us \citep{be39,bragaglia6791}, five OCs in the
Gaia-ESO Survey \citep{gilmore12}, one in APOGEE \citep{apogee}, and four more
from different groups. While not exhaustive, the OC list comprises only
clusters in which large samples of stars were analysed; for more data on OCs and
MPs, see \cite{maclean15,krause16}.

\begin{figure*}
\centering
\includegraphics[scale=0.7]{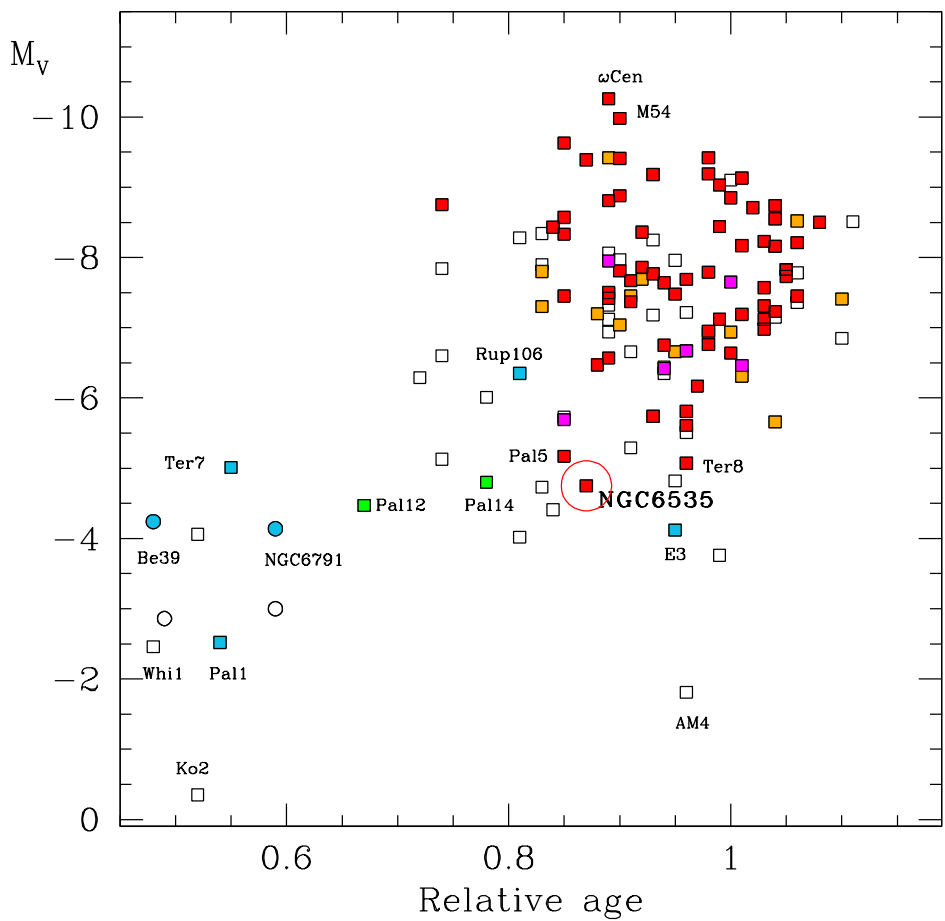}   
\caption{Relative age and absolute $V$ mag for the GCs presently in the MW 
(open squares), including those associated with the Sgr dSph and for old open
clusters (circles). Coloured symbols indicate clusters for which a) there is
positive  indication of multiple population from high-resolution spectroscopy
(red) or from photometry or low-resolution spectroscopy (orange); b) there are
uncertainties, but the presence is probable (magenta) or unlikely (green); and c)
a negative answer has been found (light blue). A few interesting objects are
labelled. } 
\label{figmvage}
\end{figure*}

The result of this process is presented in Table~\ref{tableall} and
Fig.~\ref{figmvage} for the MW clusters. The table lists all the MW GCs (plus
several OCs) for which a statement on MPs can be made. The Table  also presents the clusters'
metallicity and M$_V$ \citep[both from][]{harris}, relative age
\citep[collected from][or individual papers]{zcorri}, mass
\citep[from][whenever present]{mlvdm05}, a reference to the first paper(s)
discussing MPs or to our homogeneous survey, if available, and a flag
indicating whether MPs are actually present.  We gathered information on
90 GCs out of 157 in \cite{harris}, i.e. about 57\% of the known MW GCs. Of
these clusters, 77 show definitely the presence of MPs, 63 on the basis of Na,O and/or
Mg,Al anti-correlations, and 14 on the basis of C,N variations. For eight more
clusters the answer is uncertain, but generally the positive option is
favoured. Finally, only a handful of cases seem to host SSPs (but note Pal~12,
for which two answers are provided by different methods); all of these have a
low $M_V$, except  Ruprecht~106. On the contrary, all the OCs are SSPs.

\begin{figure}
\centering
\includegraphics[scale=0.44]{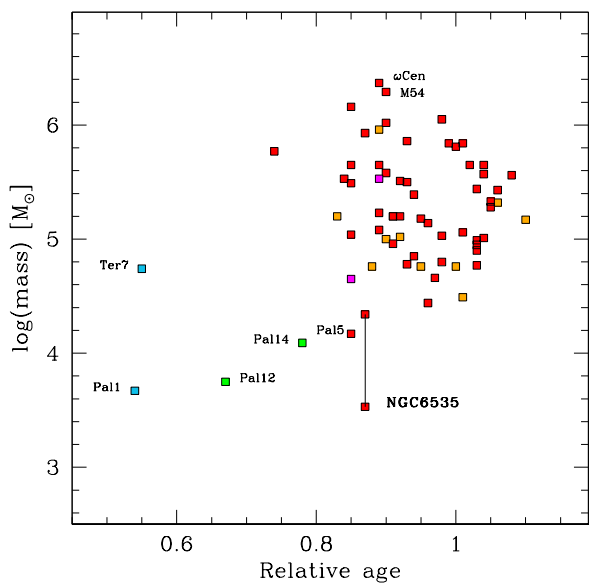}
\caption{As in the previous figure, but only for GCs and using mass instead of
M$_V$. Shown are the two values for NGC~6535 (\citealt{mlvdm05} and that
derived in Sec.\ref{kin}),  joined by a line.}
\label{figmassage}
\end{figure}

Figure~\ref{figmvage} shows graphically the same information, using M$_V$ and
relative age (in a scale where 1=13.5 Gyr). The plot confirms that NGC~6535 is
the smallest GC showing MPs and that the bulk of old, high-mass GCs hosts MPs.
Figure~\ref{figmassage} is similar, but shows the mass, taken
from \cite{mlvdm05}, for homogeneity sake; in fact, there are the two different
values for NGC~6535. Even if the information is available for  fewer GCs, the
picture is the same: cluster mass is surely one of the main drivers for the
presence/absence of MPs.

\begin{figure}
\centering
\includegraphics[scale=0.44]{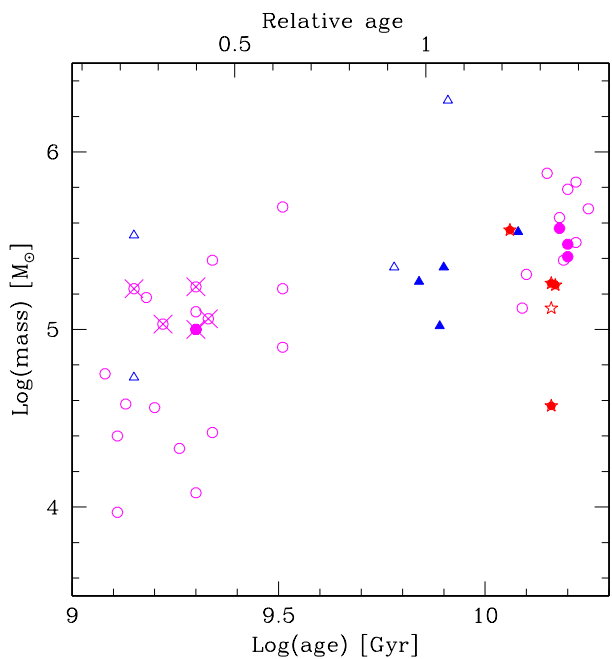}
\caption{Mass and age (in logarithm) for clusters in the LMC (magenta), SMC
(blue), and Fnx dSph (red). Open symbols indicate clusters for which we do not
have a definite answer on the presence of  MPs, crosses those
for which MPs have been excluded spectroscopically, and filled symbols those for
which MPs have been found spectroscopically and/or photometrically.}
\label{fignonmw}
\end{figure}

However, a second  factor seems important, the cluster age: with the exception
of E3, all other SSP clusters are young.  To constrain models for cluster
formation and test pollution of the SG by pristine stars, it is fundamental to  study the presence of MPs as a function of both  mass  and age. 
Of course, we must take into account that age and mass seem to be related, at least
in the MW, since massive clusters tend to be old (see
Figs.~\ref{figmvage} and \ref{figmassage}). This means that we need to resort to
extragalactic clusters, where we find young(er) and massive clusters, at
variance with the MW. By limiting the study to clusters in which individual stars can be
resolved and observed,  some clusters in nearby dwarfs have been studied. We
represent the  collected information on the extragalactic clusters in
Table~\ref{tablenonmw}, where we took metallicity from the individual papers,
mass, and age from \cite{mackey_gilmore03a,mackey_gilmore03b,mackey_gilmore03c}.
Figure~\ref{fignonmw} plots age versus mass for these LMC, SMC, and Fnx dSph
clusters.

\cite{letarte06} obtained high-resolution spectra of three stars in each of
three metal-poor GCs in Fnx dSph, finding MPs. \cite{larsen14} confirmed the
finding by means of HST near-UV and optical photometry, thereby adding a fourth
cluster. \cite{mucciarelli09} found MPs in three old GCs in the LMC on the
basis of their Na, O distribution. In all these cases the SG does not seem to
be as predominant as in the MW GCs of similar mass and age. When however
younger ages are considered, 
\cite{mucciarelli08,mucciarelli11,mucciarelli1806} did not find MPs in
intermediate-age LMC clusters, seemingly reproducing what we see in the MW (but
see below).

\cite{dalessandro16} discovered the presence of MPs in NGC~121, the only SMC
cluster as old as the classical MW GCs, using mostly near-UV HST photometry.
This was the first SMC cluster to display the presence of MPs; the discovery was confirmed by
\citealt{niederhofer17a} using a similar HST filter set. Interestingly,
\cite{dalessandro16} found a dominant FG (about 65\%), which could be due
either to a smaller formation of SG stars or to a smaller loss of FG stars than
in MW GCs of similar mass and metallicity. These authors also acquired high-resolution
spectra of five giants, but they seem to belong only to FG, according to their
Na, O and Al, Mg abundances. This may be due to small number statistics,
especially in the presence of a dominant FG, or to the decoupling of ``C,N" and
``Na,O" effects. The latter effect was touched upon in \cite{heidi} and we plan
to discuss the issue in a forthcoming paper.

Other MC clusters have indeed been found to host MPs on the basis of their
(UV-visual) CMDs and/or CN, CH band strengths; both methods rely ultimately on
the increased N and decreased C abundance in SG stars. Apart from the
increasing number of studied clusters, the interesting part is that MPs are
found in younger and younger clusters. For instance, \cite{hollyhead17}
detected MPs in the SMC cluster Lindsay~1 (age about 8 Gyr) using
low-resolution spectra and CN, CH bands; \cite{niederhofer17b}, using
photometry, confirmed the result and added two more not-so-old SMC clusters,
NGC~339 and NGC~416 (ages between 6 and 7.5 Gyr). These authors also measured the
fraction of SG in the three clusters, which always resulted to be lower than the
average MW value, at least to the values based on the Na, O anti-correlation (36\%,
25\%, and 45\% in Lindsay~1, NGC~339, and NGC~416, respectively). These
clusters are massive, about $10^5$ M$_\odot$, and are filling the gap between
the classical old MW clusters and the intermediate-age and young SMC, LMC
clusters, where no indication of MPs has been found; see
\cite{mucciarelli09,mucciarelli11,mucciarelli1806}, \cite{martocchia17}, and
Table~\ref{tablenonmw}. However, the situation evolved recently, with the
discovery of split sequences in the LMC cluster NGC~1978 (Lardo, priv. comm.;
\citealt{martocchiasubm}), which is only 2 Gyr old (translating to a redshift
$z\sim 0.15$). This cluster had been considered a SSP by
\cite{mucciarelli08}, looking at Na and O.

The presence of MPs also at young ages (i.e. in clusters formed at redshifts 
well past the epoch of GC formation) seems at odds with what we find in the MW
(see Table~\ref{tableall} and Figs.~\ref{figmvage},\ref{figmassage}) and needs
to be explained. More observations of the same clusters using both
photometry/CN bands and high-resolution spectroscopy  are also required to
clarify if they are really tracing the same phenomenon.

\section{Summary and conclusions}\label{end}

As part of our large, homogeneous survey of MW GCs with FLAMES, we observed
NGC~6535, which is possibly the lowest mass cluster in which MPs are present (see
Fig.~\ref{figmvage}). We collected our data just before the first CMDs of the
HST {\em UV Legacy Survey} were made public and we may confirm
spectroscopically that stars belonging to different stellar populations
co-exist in this cluster.

We measured abundances of Fe and other species, but concentrated here only on
the light elements. The extension of the Na-O anti-correlation has been
measured, with IQR[O/Na]=0.44, in line with the idea that the cluster has
suffered a strong mass loss. Using the Na, O distribution, we found that FG and
SG are about 30\% and 70\%, respectively, while the separation based on
UV-optical filters is close to 50-50\% \citep{milone17}. This difference is not
unusual when comparing fractions based essentially on O and Na or  C and N,
respectively. Using the RVs and King models, we estimated the cluster mass;
even if our result is larger than in \cite{mlvdm05}, we confirm that NGC~6535
has a low present-day mass.

To put NGC~6535 in context, we collected all available literature information
on the presence (or absence) of MPs in clusters. Using either M$_V$ or mass, we
confirm that this cluster lies near the lower envelope of MW GCs hosting MPs
(see  Figs.~\ref{figmvage}, \ref{figmassage}). In addition to mass, age also seems to
play a role in deciding upon the appearance of MPs and this puts important
constraints to models of cluster formation. While all young MW clusters (open
clusters) studied so far seem to be SSPs, some younger extragalactic clusters
have been found to host MPs. To really advance in our comprehension of the MP
phenomenon, further observations are required and a strong synergy between
theory/models and photometry, low-resolution, and high-resolution spectroscopy
is to be encouraged.

\begin{table*}
\centering
\caption{Information on MW GCs and OCs for which multiple populations have been
studied.}
\begin{tabular}{lccccll}
\hline
Cluster &[Fe/H] &M$_V$ &Age &LogM &MP? &Ref for MPs \\
\hline
NGC104  &-0.72  &-9.42   &0.98  &  6.05  & Y  &1,2 \\
NGC288  &-1.32  &-6.75   &0.94  &  4.85  & Y  &1,2 \\
NGC362  &-1.26  &-8.43   &0.84  &  5.53  & Y  &3   \\
NGC1851 &-1.18  &-8.33   &0.85  &  5.49  & Y  &4,5 \\
NGC1904 &-1.60  &-7.86   &0.92  &  5.20  & Y  &1,2 \\
NGC2808 &-1.14  &-9.39   &0.87  &  5.93  & Y  &6   \\
NGC3201 &-1.59  &-7.45   &0.85  &  5.04  & Y  &1,2 \\
NGC4147 &-1.80  &-6.17   &0.97  &  4.66  & Y  &7   \\
NGC4372 &-2.17  &-7.79   &0.98  &        & Y  &8   \\
NGC4590 &-2.23  &-7.37   &0.91  &  4.96  & Y  &1,2 \\
NGC4833 &-1.85  &-8.17   &1.01  &        & Y  &9   \\
NGC5024 &-2.10  &-8.71   &1.02  &  5.65  & Y  &10,11\\
NGC5053 &-2.27  &-6.76   &0.98  &  4.80  & Y  &12   \\
NGC5139 &-1.53  &-10.26  &0.89  &  6.37  & Y  &13,14  \\
NGC5272 &-1.50  &-8.88   &0.90  &  5.58  & Y  &15,16,17  \\
NGC5286 &-1.69  &-8.74   &1.04  &  5.65  & Y  &18    \\
Pal5    &-1.41  &-5.17   &0.85  &  4.17  & Y  &19    \\
NGC5466 &-1.98  &-6.98   &1.03  &  4.77  & Y  &10    \\
NGC5634 &-1.88  &-7.69   &0.96  &  5.14  & Y  &20    \\
NGC5694 &-1.98  &-7.83   &1.05  &  5.33  & Y  &21    \\
NGC5824 &-1.91  &-8.85   &1.00  &  5.81  & Y  &22    \\
NGC5927 &-0.49  &-7.81   &0.90  &        & Y  &23    \\
NGC5897 &-1.90  &-7.23   &1.04  &  5.01  & Y  &24    \\
NGC5904 &-1.29  &-8.81   &0.89  &  5.65  & Y  &1,2   \\
NGC5986 &-1.59  &-8.44   &0.99  &        & Y  &25    \\
NGC6093 &-1.75  &-8.23   &1.03  &  5.44  & Y  &26    \\
NGC6121 &-1.16  &-7.19   &1.01  &  5.06  & Y  &1,2   \\
NGC6139 &-1.65  &-8.36   &0.92  &  5.51  & Y  &27    \\
NGC6171 &-1.02  &-7.12   &1.03  &  4.90  & Y  &1,2   \\
NGC6205 &-1.53  &-8.55   &1.04  &  5.57  & Y  &16,17 \\
NGC6218 &-1.37  &-7.31   &1.03  &  4.95  & Y  &28    \\
NGC6254 &-1.56  &-7.48   &0.95  &  5.18  & Y  &1,2   \\
NGC6266 &-1.18  &-9.18   &0.93  &  5.86  & Y  &29,30 \\
NGC6273 &-1.74  &-9.13   &1.01  &  5.84  & Y  &31    \\
NGC6341 &-2.31  &-8.21   &1.06  &  5.43  & Y  &10    \\
NGC6352 &-0.64  &-6.47   &0.88  &        & Y  &32    \\
NGC6362 &-0.99  &-6.95   &0.98  &  5.03  & Y  &33    \\
NGC6366 &-0.59  &-5.74   &0.93  &  4.78  & Y  &34    \\
NGC6388 &-0.55  &-9.41   &0.90  &  6.02  & Y  &35,1  \\
NGC6397 &-2.02  &-6.64   &1.00  &        & Y  &1,2   \\
NGC6440 &-0.36  &-8.75   &0.74  &  5.77  & Y  &36    \\
NGC6441 &-0.46  &-9.63   &0.85  &  6.16  & Y  &37,38 \\
NGC6528 &-0.11  &-6.57   &0.89  &  5.08  & Y  &39    \\
{\bf NGC6535} &{\bf -1.79}      &{\bf -4.75}     &{\bf 0.87}  &  {\bf 3.53}  & {\bf Y}  &{\bf This paper}                   \\
NGC6553 &-0.18  &-7.77   &0.93  &  5.50  & Y  &39,40  \\
NGC6626 &-1.32  &-8.16   &1.04  &        & Y  &41     \\
NGC6637 &-0.64  &-7.64   &0.94  &  5.39  & Y  &42     \\
NGC6656 &-1.70  &-8.50   &1.08  &  5.56  & Y  &43,44  \\
NGC6681 &-1.62  &-7.12   &0.99  &        & Y  &45,46  \\
NGC6712 &-1.02  &-7.50   &0.89  &  5.23  & Y  &47     \\
NGC6715 &-1.49  &-9.98   &0.90  &  6.29  & Y  &48,49  \\
NGC6723 &-1.10  &-7.83   &1.05  &  5.28  & Y  &50     \\
NGC6752 &-1.54  &-7.73   &1.05  &        & Y  &51     \\
NGC6809 &-1.94  &-7.57   &1.03  &  4.99  & Y  &1,2    \\
NGC6838 &-0.78  &-5.61   &0.96  &        & Y  &1,2    \\
NGC6864 &-1.29  &-8.57   &0.85  &  5.65  & Y  &52     \\
Terzan8 &-2.16  &-5.07   &0.96  &        & Y  &53     \\
NGC7006 &-1.52  &-7.67   &0.91  &  5.20  & Y  &54     \\
NGC7078 &-2.37  &-9.19   &0.98  &        & Y  &1,2    \\
NGC7089 &-1.65  &-9.03   &0.99  &  5.84  & Y  &55     \\
NGC7099 &-2.27  &-7.45   &1.06  &        & Y  &1,2    \\
NGC7492 &-1.78  &-5.81   &0.96  &  4.44  & Y  &56     \\
Terzan5 &-0.23  &-7.42   &0.89  &        & Y  &39     \\
\hline
\end{tabular}
\label{tableall}
\end{table*}

\begin{table*}
\centering
\setcounter{table}{8}
\caption{(continued)}
\begin{tabular}{lccccll}
\hline
Cluster &[Fe/H] &M$_V$ &Age &LogM &MP? &Ref for MPs \\
\hline
NGC1261 &-1.27  &-7.80   &0.83  &  5.20  & P  &46,57  \\
NGC2298 &-1.92  &-6.31   &1.01  &  4.49  & P  &46,57  \\
NGC2419 &-2.15  &-9.42   &0.89  &  5.96  & P  &58,59,60,61\\   
NGC6101 &-1.98  &-6.94   &1.00  &  4.76  & P  &46,57  \\
NGC6304 &-0.45  &-7.30   &0.83  &        & P  &46,57  \\
NGC6496 &-0.46  &-7.20   &0.88  &  4.76  & P  &46,57  \\
NGC6541 &-1.81  &-8.52   &1.06  &  5.32  & P  &62,46,57 \\
NGC6584 &-1.50  &-7.69   &0.92  &  5.02  & P  &46,57  \\
NGC6624 &-0.44  &-7.49   &0.89  &        & P  &46,57  \\
NGC6652 &-0.81  &-6.66   &0.95  &  4.76  & P  &46,57  \\
NGC6717 &-1.26  &-5.66   &1.04  &        & P  &46,57  \\
NGC6779 &-1.98  &-7.41   &1.10  &  5.17  & P  &46,57  \\
NGC6934 &-1.47  &-7.45   &0.91  &  5.20  & P  &63,46,57 \\
NGC6981 &-1.42  &-7.04   &0.90  &  5.00  & P  &46,57 \\
Pal6    &-0.91  &-6.79   &      &        & Y? &39    \\
NGC6333 &-1.77  &-7.95   &0.89  &  5.53  & Y? &64    \\
NGC6426 &-2.15  &-6.67   &0.96  &        & Y? &65    \\
NGC6522 &-1.34  &-7.65   &1.00  &        & Y? &39    \\
NGC6342 &-0.55  &-6.42   &0.94  &        & Y? &34    \\
HP1     &-1.00  &-6.46   &1.01  &        & Y? &66    \\
Pal3    &-1.63  &-5.69   &0.85  &  4.65  & Y? &67    \\
Pal14   &-1.62  &-4.80   &0.78  &  4.09  & N? &68    \\
Pal12   &-0.85  &-4.47   &0.67  &  3.75  & N/P&69,70 \\
E3      &-0.83  &-4.12   &0.95  &        & N  &71    \\
Terzan7 &-0.32  &-5.01   &0.55  &  4.74  & N  &72,73 \\
Rup106  &-1.68  &-6.35   &0.81  &        & N  &74    \\
Pal1    &-0.65  &-2.52   &0.54  &  3.67  & N  & 75   \\
\hline
NGC6791     &+0.40 &-4.14 &0.59 & & N &76,77,78 \\
Berkeley39  &-0.20 &-4.24 &0.48 & & N &79  \\
Trumpler20  &+0.17 &      &0.11 & & N &80  \\
NGC6705     &+0.10 &-6.00 &0.02 & & N &81  \\
Berkeley81  &+0.23 &-4.70 &0.07 & & N &82  \\
Trumpler23  &+0.14 &      &0.06 & & N &83  \\
NGC6802     &+0.10 &      &0.07 & & N &84  \\
NGC2420     &-0.16 &-3.44 &0.15 & & N &85  \\
NGC7789     &+0.03 &      &0.12 & & N &86  \\
Collinder261&-0.03 &-2.87 &0.44 & & N &87  \\
NGC752      &-0.02 &      &0.12 & & N &88  \\
NGC2682     &+0.03 &-3.16 &0.30 & & N &89  \\
\hline
\end{tabular}
\label{tableall}
\tablefoot{
Metallicity and M$_V$ for GCs come from \cite{harris}; Age=relative age come
from \cite{zcorri} or from individual papers; LogM=Log(mass)come from
\cite{mlvdm05}. The reference for presence of multiple stellar populations is
to our homogeneous FLAMES survey or to the first paper(s) presenting evidence
for this survey. The codes for multiple populations are Y=evidence from high-resolution
spectroscopy (i.e. O,Na; Mg,Al); Y?=dubious but probable; N?=dubious but
improbable; N=not present; P=present, on the basis of low-resolution
spectroscopy or photometric data (i.e. CN,CH; split/spread sequences).\\
Values for M$_V$ for OCs come from \cite{lata02}, relative age and metallicity
from the individual  papers. $APO$ stands for APOGEE, $GES$ for Gaia-ESO
survey.\\
References.
(1)  \cite{carretta09a}; 
(2)  \cite{carretta09b};    
(3)  \cite{n362};                 
(4)  \cite{n1851a};
(5)  \cite{n1851};        
(6)  \cite{carretta06};           
(7)  \cite{villanova4147};        
(8)  \cite{sanroman15} ;          
(9)  \cite{n4833};                
(10) \cite{meszaros15};
(11) \cite{boberg16};
(12) \cite{boberg15};     
(13) \cite{jonsonomega};
(14) \cite{marinoomega};
(15) \cite{cohen78};
(16) \cite{kraft92};
(17) \cite{m3_m13};
(18) \cite{marino15};   
(19) \cite{smithpal15}; 
(20) \cite{n5634};              
(21) \cite{mucciarelli13};      
(22) \cite{roederer16};         
(23) \cite{pancinoges};         
(24) \cite{koch_mcwilliam};     
(25) \cite{johnson17};               
(26) \cite{m80};        
(27) \cite{bragaglia6139};      
(28) \cite{n6218};              
(29) \cite{yong14a};
(30) \cite{lapenna15};  
(31) \cite{johnson15};          
(32) \cite{feltzing09};         
(33) \cite{mucciarelli16};      
(34) \cite{johnson16};  
(35) \cite{n6388};   
(36) \cite{munoz17};
(37) \cite{gratton06};
(38) \cite{gratton07}; 
(39) \cite{schiavon17};   
(40) \cite{tang17a};
(41) \cite{villanova17}; 
(42) \cite{lee07};  
(43) \cite{cohenm22};
(44) \cite{marinom22}; 
(45) \cite{omalley17};
(46) \cite{soto17};   
(47) \cite{yong08};               
(48) \cite{m54a};
(49) \cite{m54b};         
(50) \cite{gratton15};          
(51) \cite{n6752};      
(52) \cite{kacharov14};         
(53) \cite{ter8};  
(54) \cite{kraft98};
(55) \cite{yong14b};            
(56) \cite{cohen_melendez};     
(57) \cite{milone17};           
(58) \cite{cohen_kirby}
(59) \cite{mucciarelli12};
(60) \cite{beccari13};
(61) \cite{frank15}; 
(62) \cite{geisler88};            
(63) \cite{smith_bell86};         
(64) \cite{johnson13};  
(65) \cite{hanke17};                      
(66) \cite{barbuy16};                   
(67) \cite{koch09};             
(68) \cite{caliskan12};                 
(69) \cite{cohenpal12};
(70) \cite{pancino10};                          
(71) \cite{salinas_strader};            
(72) \cite{taut04};
(73) \cite{sbordoneter7}; 
(74) \cite{villanovarup106};                    
(75) \cite{sakari11};                   
(76) \cite{bragaglia6791};
(77) \cite{cunha15} ($APO$);
(78) \cite{geisler12};
(79) \cite{be39}; 
(80) \cite{donati14} ($GES$);
(81) \cite{m11} ($GES$); 
(82) \cite{magrini15} ($GES$); 
(83) \cite{overbeek17}  ($GES$);
(84) \cite{tang17b} ($GES$);
(85) \cite{souto16} ($APO$);
(86) \cite{overbeek15};
(87) \cite{desilva07};
(88) \cite{bocektopcu15};
(89) \cite{randich06}.}
\end{table*}

\begin{table*}
\centering
\caption{Information on non-MW GCs for which multiple populations have been
studied
}
\begin{tabular}{lccccll}
\hline
Cluster &[Fe/H] &LogMass &LogAge &MP? &Ref for MPs \\
\hline
LMC NGC1786 &-1.87 &5.57 &10.18 &Y   &1   \\
LMC NGC2210 &-1.97 &5.48 &10.20 &Y   &1   \\
LMC NGC2257 &-1.63 &5.41 &10.20 &Y   &1   \\
SMC NGC121  &-1.71 &5.55 &10.08 &Y   &2,3 \\
FNX Fnx1    &-2.5  &4.57 &10.16 &Y   &4,5 \\
FNX Fnx2    &-2.1  &5.26 &10.16 &Y   &4,5 \\
FNX Fnx3    &-2.3  &5.56 &10.06 &Y   &4,5 \\
FNX Fnx5    &-2.1  &5.25 &10.17 &P   &5   \\
SMC Lindsay1&-1.35 &5.35 & 9.90 &P   &6,7 \\
SMC NGC339  &-1.50 &5.02 & 9.89 &P   &7   \\
SMC NGC416  &-1.41 &5.27 & 9.84 &P   &7   \\
SMC NGC419  &      &     & 9.18 &N   &8   \\
LMC NGC1978 &      &5    & 9.3  &N,P &9,10\\
LMC NGC1651 &-0.37 &5.24 & 9.30 &N   &9   \\
LMC NGC1806 &-0.6  &5.03 & 9.22 &N   &11  \\
LMC NGC1783 &      &     &      &N   &9   \\
LMC NGC1866 &-0.50 &4.63 & 8.12 &N   &12  \\
LMC NGC2173 &-0.24 &5.06 & 9.33 &N   &9   \\
\hline
\end{tabular}
\label{tablenonmw}
\tablefoot{Mass and age taken from
\cite{mackey_gilmore03a,mackey_gilmore03b,mackey_gilmore03c} or from individual
papers; metallicity from individual papers. The flag for multiple population
has the same code as the previous table. \\
References. 
(1) \cite{mucciarelli09};
(2) \cite{dalessandro16}; 
(3) P: \cite{niederhofer17a};
(4) \cite{letarte06}; 
(5) P: \citealt{larsen14};
(6) CN: \cite{hollyhead17}; 
(7) P: \cite{niederhofer17b};
(8)       \cite{martocchia17} (phot.);
(9) N: \cite{mucciarelli08}; 
(10) P: \cite{martocchiasubm};
(11)      \cite{mucciarelli1806};
(12)      \cite{mucciarelli11};
}  
\end{table*}

\begin{acknowledgements}
This research has made use of Vizier and SIMBAD, operated at CDS, Strasbourg,
France,  NASA's Astrophysical Data System, and {\sc TOPCAT}
(http://www.starlink.ac.uk/topcat/). The Guide Star Catalogue-II is a joint
project of the Space Telescope Science Institute and the Osservatorio
Astronomico di Torino. Space Telescope Science Institute is operated by the
Association of Universities for Research in Astronomy, for the National
Aeronautics and Space Administration under contract NAS5-26555. The
participation of the Osservatorio Astronomico di Torino is supported by the
Italian Council for Research in Astronomy. Additional support is provided by
European Southern Observatory, Space Telescope European Coordinating Facility,
the International GEMINI project, and the European Space Agency Astrophysics
Division. This publication makes use of data products from the Two Micron All
Sky Survey, which is a joint project of the University of Massachusetts and the
Infrared Processing and Analysis Center/California Institute of Technology,
funded by the National Aeronautics and Space Administration and the National
Science Foundation. Some of the data presented in this paper were obtained from
the Mikulski Archive for Space Telescopes (MAST). STScI is operated by the
Association of Universities for Research in Astronomy, Inc., under NASA
contract NAS5-26555. Support for MAST for non-HST data is provided by the NASA
Office of Space Science via grant NNX09AF08G and by other grants and
contracts. 
\end{acknowledgements}

\end{document}